\documentclass[a4paper,fleqn,usenatbib]{mnras}

\usepackage[T1]{fontenc}
\usepackage{ae,aecompl}

\usepackage{graphicx}	
\usepackage{amsmath}	
\usepackage{amssymb}	

\newcommand{\msun}{\,M$_{\odot}$}

\newcommand{\lsun}{\,L$_{\odot}$}
\newcommand{\kms}{\,km~s$^{-1}$}

\newcommand{\ang}{\,$\rm{\AA}$}
\newcommand{\ergs}{\,erg~s$^{-1}$}
\newcommand{\halpha}{\,H$\alpha$}
\newcommand{\hbeta}{\,H$\beta$}
\newcommand{\hgamma}{\,H$\gamma$}

\title[LSQ13zm: an outburst heralds the death of a massive star.]{Interacting supernovae and supernova impostors. LSQ13zm: an outburst heralds the death of a massive star.}

\author[L.~Tartaglia et al.]{L.~Tartaglia$^{1,2}$\thanks{E-mail: leonardo.tartaglia@oapd.inaf.it}, 
A.~Pastorello$^{2}$,
M.~Sullivan$^{3}$,
C.~Baltay$^{4}$,
D.~Rabinowitz$^{5}$,
P.~Nugent$^{5,6}$,
\newauthor 
A.J.~Drake$^{7}$,
S.G.~Djorgovski$^{7}$,
A.~Gal-Yam$^{8}$,
S.~Fabrika$^{10,11}$,
E.A.~Barsukova$^{10}$,
\newauthor
V.P.~Goranskij$^{9}$,
A.F.~Valeev$^{10,11}$,
T.~Fatkhullin$^{10}$, 
S.~Schulze$^{12,13}$, 
A.~Mehner$^{14}$, 
\newauthor 
F.E.~Bauer$^{13,12,15}$, 
S.~Taubenberger$^{16,17,}$, 
J.~Nordin$^{17}$, 
S.~Valenti$^{18,19}$, 
D.A.~Howell$^{18,19}$, 
\newauthor
S.~Benetti$^{1}$, 
E.~Cappellaro$^{1}$, 
G.~Fasano$^{1}$, 
N.~Elias-Rosa$^{1}$, 
M.~Barbieri$^{20}$, 
D.~Bettoni$^{1}$, 
\newauthor
A.~Harutyunyan$^{21}$, 
T.~Kangas$^{22}$,
E.~Kankare$^{23}$,
J.C.~Martin$^{24}$, 
S.~Mattila$^{25,22}$,
\newauthor 
A~Morales-Garoffolo$^{26}$, 
P.~Ochner$^{1}$,
Umaa D. Rebbapragada$^{27}$, 
G.~Terreran$^{1,23}$, 
\newauthor
L.~Tomasella$^{1}$, 
M.~Turatto$^{1}$, 
E.~Verroi$^{28}$.
P.~R.~Wo\'zniak$^{29}$, 
\newauthor
{\small \it Affiliations are listed at the end of the paper}
}

\date{Accepted 2016 March 18. Received 2016 March 18; in original form 2015 September 29}
\pubyear{2016}

\begin{document}
\label{firstpage}
\pagerange{\pageref{firstpage}--\pageref{lastpage}}
\maketitle

\begin{abstract}
We report photometric and spectroscopic observations of the optical transient LSQ13zm.
Historical data reveal the presence of an eruptive episode (that we label as `2013a') followed by a much brighter outburst (`2013b') three weeks later, that we argue to be the genuine supernova explosion. 
This sequence of events closely resemble those observed for SN~2010mc and (in 2012) SN~2009ip.
The absolute magnitude reached by LSQ13zm during 2013a ($M_R=-14.87\pm0.25\,\rm{mag}$) is comparable with those of supernova impostors, while that of the 2013b event ($M_R=-18.46\pm0.21\,\rm{mag}$) is consistent with those of interacting supernovae.
Our spectra reveal the presence of a dense and structured circumstellar medium, probably produced through numerous pre-supernova mass-loss events. 
In addition, we find evidence for high-velocity ejecta, with a fraction of gas expelled at more than 20000\kms.
The spectra of LSQ13zm show remarkable similarity with those of well-studied core-collapse supernovae.
From the analysis of the available photometric and spectroscopic data, we conclude that we first observed the last event of an eruptive sequence from a massive star, likely a Luminous Blue Variable, which a short time later exploded as a core-collapse supernova.
The detailed analysis of archival images suggest that the host galaxy is a star-forming Blue Dwarf Compact Galaxy.
\end{abstract}


%%%%%%%%
% SECTION 1
%%%%%%%%
\section{Introduction} \label{intro}
It is widely accepted that most interacting supernovae (SNe) arise from the collapse of the nucleus of massive stars ($M\gtrsim8$\msun) exploding in a dense circumstellar medium (CSM), although a fraction of them could result from thermonuclear explosions of white dwarfs in binary systems \citep[e.g.][]{2012Sci...337..942D}. 
In general, the cocoon-like CSM surrounding the progenitor system is generated via stationary stellar winds, binary interaction or even multiple eruptive events as a consequence of instabilities during the latest stages of the stellar life. \\

The most common interacting core-collapse SNe (CCSNe) have spectra dominated by H lines. 
Their spectra are characterised by a blue continuum with superimposed prominent Balmer lines in emission, usually showing profiles with multiple components.
These are probably produced in gas shells moving at different velocities, and having different temperatures and densities \citep{1993MNRAS.262..128T}. 
In particular, the spectra show narrow components in emission at all stages of the SN evolution, which are recombination lines emitted by the un-shocked photo-ionised CSM, although, in some cases, narrow components disappear soon after maximum, as the un-shocked CSM recombines \citep[see e.g. SN~1998S;][]{2001MNRAS.325..907F}
This gas is located in the outer circumstellar environment, and moves at relatively low velocities, from a few tens to about one thousand\kms~\citep[see e.g.][]{2012ApJ...744...10K}.
Because of the presence of these narrow spectral lines, H-rich interacting SNe are labelled as `Type IIn' SNe \citep{1990MNRAS.244..269S,1997ARA&A..35..309F}. \\

Intermediate components in the line profiles (with inferred velocities of a few 1000\kms) are also frequently observed, and are considered as one of the signatures of interaction between the SN ejecta and the CSM, since they probably form in the regions of shocked gas \citep{1994MNRAS.268..173C}.
It is important to remark that the photo-ionised un-shocked CSM and the shocked gas interface frequently mask the freely expanding SN ejecta \citep{2002ApJ...572..350F}, limiting our understanding of the explosion mechanism.
However, when the CSM is optically thin or because of its particular geometry, the broad lines associated with the SN ejecta can be observed.
The presence of high-velocity ejecta ($\gtrsim10^4$\kms~for the bulk of the material) along with the high temperatures of the ejected gas ($\gtrsim10^4$~K, usually inferred from a black-body fit to the spectral continuum), and the slow colour/temperature evolution are key ingredients to characterise the explosion. \\

The photometric evolution of SNe IIn is usually slow \citep[although in some cases, fast declines are observed;][]{2000MNRAS.318.1093F,2001MNRAS.325..907F,2002ApJ...573..144D}, with luminous light curve peaks \citep[absolute magnitudes ranging from $-18$ to $-22$;][]{2012ApJ...756..173S,2012ApJ...744...10K}.
When at later phases the SN luminosity is still dominated by CSM-ejecta interaction, the light curve remains more luminous than that predicted for the $^{56}$Co to $^{56}$Fe decay. \\

A key improvement in our comprehension of SNe IIn resulted from the evidence of a connection \citep{2006A&A...460L...5K,2006ApJ...645L..45S} between some members of this SN type with a rare class of very massive and unstable stars, the so-called `Luminous Blue Variables' \citep[LBVs;][]{1994PASP..106.1025H}.
LBVs are luminous ($10^5$ - $10^6$~\lsun) and very massive ($M\gtrsim30$\msun) evolved stars close to the Eddington limit, characterised by an erratic instability of their outer layers and a high rate of mass loss ($>10^{-4}$\msun~$\rm{yr}^{-1}$). 
These stars normally sit into the `S-Doradus instability strip' in the Hertzsprung-Russel (H-R) Diagram \citep[within the luminosity-temperature range $-9\geqslant M_{\rm{bol}}\geqslant-11$, $8500\,\rm{K}\leqslant T_{\rm{eff}}\leqslant35000$~K;][]{1989A&A...217...87W,2004ApJ...615..475S}. 
Moderate `S~Doradus (S~Dor) type' variability, is related to mass-loss episodes involving the outer layers.
During this phase, lasting from years to decades, the star moves to the red region of the H-R Diagram, increasing its optical luminosity by 1--2~mag, without changing significantly its bolometric luminosity \citep[although fluctuations in the bolometric luminosity has been observed during the S--Dor phase of AG~Car;][]{2009ApJ...698.1698G}. 
Micro-variations, of the order of a few tenths magnitudes on a time scale of weeks to months are also usually observed.
During their quiescent phase LBVs experience typical supergiant mass-loss rates ($\sim10^{-7}$\msun$\,\rm{yr}^{-1}$), which significantly increase during the S--Dor phase (up to $\sim10^{-5}$\msun$\,\rm{yr}^{-1}$).
Occasionally LBVs may produce giant eruptions (like that observed in $\eta$ Carinae in the mid-19th century), during which they lose a significant mass fraction of their envelope (up to $\simeq10$\msun, with mass-loss rates exceeding $\sim10^{-4}$ -- $10^{-3}$\msun$\,\rm{yr}^{-1}$) and experience a dramatic increase (3--6~mag) in luminosity \citep{1994PASP..106.1025H}, reaching absolute bolometric magnitudes at peak of above $-14\,\rm{mag}$.
The mechanism triggering the giant eruptions is not fully understood, although a few scenarios have been proposed \citep[see e.g.][]{2007Natur.450..390W,2002A&A...395L...1L}. \\

The link between some SNe IIn and LBVs is based on the detection of the progenitor stars of two type IIn SNe in archival Hubble Space Telescope (HST) images, viz. SN~2005gl \citep{2007ApJ...656..372G,2009Natur.458..865G} and SN~2010jl \citep{2011ApJ...732...63S}. 
The putative progenitors both showed absolute magnitudes consistent with those observed in quiescent LBVs. 
An indirect clue of this connection is given by signatures of a structured CSM in the spectra of some type IIn SNe, including SN~2005gj \citep{2008A&A...483L..47T}, with evidence of several shell-like layers inferred from the presence of multiple absorption features with bulk velocities consistent with those of LBV winds. \\

A further step in establishing a connection between LBVs and SNe IIn has been made through the study of the pre-SN photometric variability of the precursor stars, using archival images collected months to years before the SN explosion. 
The data archive inspection is, in fact, an invaluable tool to characterise the final stages of the progenitors of interacting transients.
Weak transient events with luminosities consistent with those expected in LBV outbursts are occasionally detected weeks to years before major re-brightenings \citep[e.g.,][]{2013ApJ...767....1P}. 
Some of them have been proposed to be sequential events leading to a SN explosion \citep{2013MNRAS.430.1801M,2013ApJ...779L...8F,2013Natur.494...65O}. 
Even more robust is the spatial coincidence between the CC SN~2006jc \citep[a Type Ibn event, see][for more details on this subclass of stripped-envelope SNe]{2000AJ....119.2303M,2008MNRAS.389..131P,2015MNRAS.449.1954P} and a stellar outburst of $\simeq-14\,\rm{mag}$ which had occurred $\sim2$~years before \citep{2007Natur.447..829P,2007ApJ...657L.105F}.
In that case, the massive precursor was likely a Wolf-Rayet (WR) star \citep{2007ApJ...657L.105F,2007Natur.447..829P,2008ApJ...687.1208T} with a residual LBV-like instability. \\

Nonetheless, in most cases, outbursts attributed to extragalactic massive stars are registered as isolated events \citep[see e.g.][for a detailed analysis on the eruption frequencies for these objects]{2014ApJ...789..104O}.
These are fainter than real SNe ($M\simeq-12$ to $-14\,\rm{mag}$), but mimic the behavior of Type IIn SNe, showing similar spectra dominated by prominent narrow Balmer lines in emission, and sometimes even similar light-curves.
They are usually labelled as `supernova impostors' \citep{2000PASP..112.1532V}, as they are not terminal SN explosions. 
Accounting for the evidence of photometric variability from massive stars, a sequential event chain linking LBVs, SN impostors and SNe IIn has been proposed \citep{2006A&A...460L...5K}. \\

In this context, it is worth mentioning the controversial case of SN~2009ip, an interacting transient whose real nature (SN explosion vs. non-terminal outburst) is still debated\footnote{Different interpretations have been proposed by \citet{2013ApJ...767....1P,2013MNRAS.433.1312F,2013MNRAS.434.2721S,2013ApJ...768...47O,2013ApJ...763L..27P,2013ApJ...764L...6S,2014ApJ...780...21M,2014MNRAS.438.1191S,2014MNRAS.442.1166M,2014ApJ...787..163G,2014AJ....147...23L,2015ApJ...803L..26M,2015AJ....149....9M,2015MNRAS.453.3886F}}.
The detection of the progenitor star in archival HST images proved it to be consistent with a massive star -- most likely an LBV -- with a zero-age-main-sequence (ZAMS) mass of $\simeq60$\msun~\citep{2010AJ....139.1451S,2011ApJ...732...32F}.
SN~2009ip was well studied in the years ahead of the putative SN explosion and exhibited erratic luminosity oscillations since summer 2009 \citep{2013ApJ...767....1P}. 
In July 2012, it experienced a further re-brightening lasting a few weeks, $\sim20$~days before a major outburst, in which the object reached an absolute magnitude competing with those of Type IIn SNe. 
However, a conclusive proof of the terminal SN explosion is still missing \citep{2015MNRAS.453.3886F}, as the expected spectral signatures (such as nucleosynthesized elements) of a SN produced in the explosion of a very massive star have not been detected yet in the spectra of SN 2009ip \citep{2013MNRAS.433.1312F}. 
A similar sequence of events was observed also for SN~2010mc \citep{2013Natur.494...65O,2014MNRAS.438.1191S}, with the detection of an outburst $\sim40$~days before the putative terminal SN explosion. \\

SN~2011ht \citep{2013ApJ...779L...8F,2013MNRAS.431.2599M} is another interesting example.
It was classified as a SN impostor \citep{2011CBET.2851....2P}, before showing a significant spectral metamorphosis which led \citet{2011CBET.2903....1P} to suggest its re-classification as a Type IIn SN.
\citet{2013ApJ...779L...8F} later reported the detection of an outburst $\sim1$~year prior to the SN explosion.
Nonetheless, also in this case, its nature is also not fully clarified \citep{2012ApJ...760...93H}. \\

The LBV stage is a short-duration phase in the life of very massive stars {\bf ($\rm{M}\gtrsim30$\msun)}, which are then expected to become H-stripped WR stars before exploding -- after a relatively long time \citep[a few $10^5$~yr, see][]{2012A&A...542A..29G} -- as Type Ib/c SNe. However, current stellar evolution codes do not predict the explosion of a CCSN soon after a major instability phase \citep{2013A&A...550L...7G}.
Nonetheless, in the light of the sequences of events involving interacting transients that have been observed in growing number, ad hoc scenarios have been proposed \citep[see e.g.][]{2014ApJ...796..121J}, and some efforts have been devoted to include the effects of instabilities in stellar evolution codes \citep{2014ApJ...785...82S}. \\

In this context, we report the results of the follow-up campaign of LSQ13zm, observed in the galaxy SDSS~J102654.56+195254.8.
The transient was discovered by the La Silla Quest (LSQ) survey\footnote{\url{http://hep.yale.edu/lasillaquest}}, and later classified as a young Type IIn SN by the Nearby Supernova Factory \citep[SNF\footnote{\url{http://snfactory.lbl.gov/}};][]{2013ATel.4994....1B} using the SuperNova Integral Field Spectrograph \citep[SNIFS,][]{2002SPIE.4836...61A} mounted on the University of Hawaii 2.2~m telescope. 
Archival data from different surveys, namely LSQ, the Intermediate Palomar Transient Factory \citep[iPTF\footnote{\url{http://www.ptf.caltech.edu/iptf}};][where the transient was designated as iPTF13ajw]{2009PASP..121.1395L,2009PASP..121.1334R} and the Catalina Real-Time Transient Survey \citep[CRTS\footnote{\url{http://crts.caltech.edu/}};][]{2009ApJ...696..870D,djorgovski2012} revealed an outburst (reaching an absolute magnitude at peak of $M_R=-14.87\pm0.25\,\rm{mag}$) $\sim3$~weeks before a major re-brightening, where the object reached an absolute magnitude $M_R=-18.46\pm0.21\,\rm{mag}$ (see Section~\ref{photoanalysis}).
Hereafter, we will refer to the first outburst as the `2013a event' and the second, more luminous re-brightening as the `2013b' event, in analogy to what has been proposed for SN~2009ip \citep{2013ApJ...767....1P}.\\

This paper is organised as follows.
In Section~\ref{host}, we characterise the galaxy hosting LSQ13zm. 
Sections~\ref{photometry}~and~\ref{spectroscopy} report the results of our photometric and spectroscopic follow-up campaigns, highlighting a few observational features, while in Section~\ref{preSN} the historical data are discussed in order to constrain the pre-SN behavior of the progenitor. 
The results are discussed and used in Section~\ref{discussion} in order to characterise the explosion scenario. 
Section~\ref{conclusions} summarises the main results of our study.

%%%%%%%%
% SECTION 2
%%%%%%%%
\section{The host galaxy} \label{host}
\begin{figure}
\includegraphics[width=1.0\columnwidth]{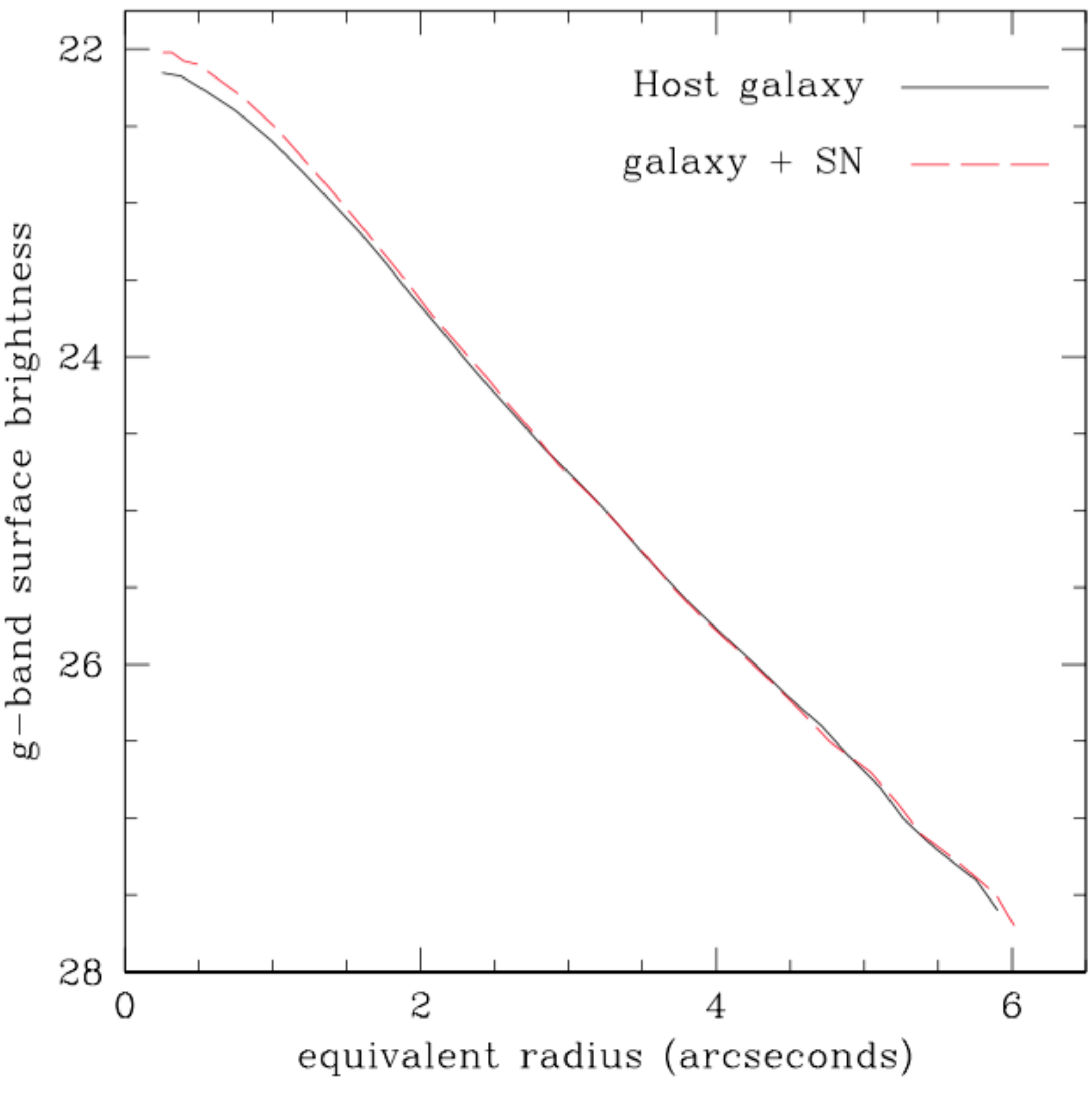}
\caption{$g$-band surface brightness radial profiles of SDSS~J102654.56$+$195254.8. The red dashed profile was obtained from a late phase image ($+284$~d) while the black one was obtained from our template image. Both images were obtained with the GTC.}
\label{radial_profiles}
\end{figure}

\begin{figure} 
\includegraphics[width=1.0\columnwidth]{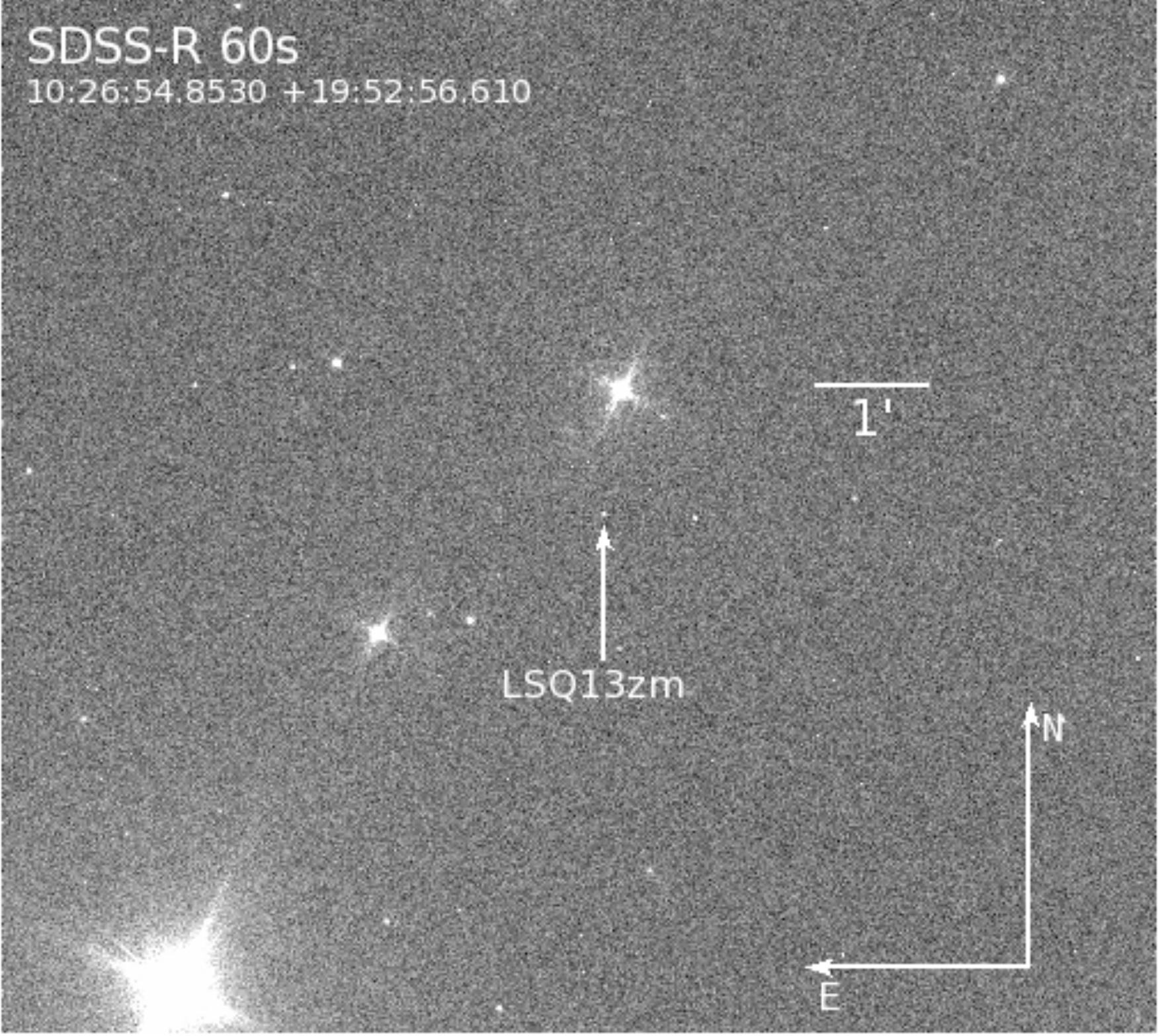}
\caption{Finding chart of LSQ13zm. Information about the orientation, scale, filter and exposure time are reported.}
\label{fc}
\end{figure}

SDSS~J102654.56$+$195254.8, the host galaxy of LSQ13zm, is a dwarf galaxy with an apparent $g$-band magnitude (as reported in the SDSS archive) of $18.95\pm0.02\,\rm{mag}$. 
Figure~\ref{radial_profiles} shows the radial profile of the surface brightness of the host computed at different epochs, and suggests that the position of the SN (RA$=$10:26:54.591 and Dec$=+$19:52:54.91~[J2000]), accurately pin-pointed from a late-phase $g$-band template-subtracted image, is almost coincident with the coordinates of the host galaxy nucleus.
% Figure 1
A finding chart of LSQ13zm is shown in Figure~\ref{fc}.
The field containing SDSS~J102654.56$+$195254.8 was observed by the Sloan Digital Sky Survey (SDSS\footnote{\url{http://www.sdss.org/}}) on 2005 March 10, and these data were used as template images for our $griz$ early-phase photometric data. 
No source is visible at the galaxy position in the 2 Micron All Sky Survey (2MASS\footnote{\url{http://www.ipac.caltech.edu/2mass/}}), while the Wide-field Infrared Survey Explorer (WISE\footnote{\url{http://www.nasa.gov/mission\_pages/WISE/main/index.html}}) catalog reports the following magnitudes: W$_1=17.006\pm0.107$, W$_2>17.321$, W$_3>12.778$, suggesting a steep decline in the spectral energy distribution (SED) of the host from the optical to the infrared (IR) bands.
From the average positions of the Balmer emission lines \halpha~and \hbeta~in the spectra of LSQ13zm (see Section~\ref{spectroscopy}), we derived a redshift of 0.029.
Adopting a standard cosmology ($H_0=73$\kms~$\rm{Mpc}^{-1}$, $\Omega_M=0.27$, $\Omega_{\Lambda}=0.73$) and using Ned Wright's Cosmological Calculator\footnote{\url{http://www.astro.ucla.edu/~wright/CosmoCalc.html}} \citep{2006PASP..118.1711W}, we derived a luminosity distance $D_L=122.0\pm8.2$~Mpc, and hence a distance modulus of $\mu=35.43\pm0.21\,\rm{mag}$.
For the foreground Galactic extinction, we adopt $A_V=0.052\,\rm{mag}$, as derived from the \cite{2011ApJ...737..103S} infrared-based dust map available through the NED\footnote{\url{http://ned.ipac.caltech.edu/}} database.
Our spectroscopic analysis, detailed in Section~\ref{spectroscopy}, reveals no evidence of narrow absorption lines of the \ion{Na}{ID} doublet at the recessional velocity of the host galaxy. 
For this reason, we will assume hereafter a negligible contribution of the host galaxy to the total extinction towards LSQ13zm.\\
% Figure 2

An accurate study of the host galaxy was performed through our template images obtained with the 10.4~m Gran Telescopio Canarias (GTC) located at the Observatorio del Roque de los Muchachos (ORM, La Palma, Canary Islands, Spain) when the SN faded below the detection threshold.
We first fit the isophotes with ellipses (1D fit), obtaining the following total magnitudes: $g=19.68\pm0.01\,\rm{mag}$, $r=19.14\pm0.02\,\rm{mag}$, $i=18.97\pm0.01\,\rm{mag}$, $z=18.73\pm0.02\,\rm{mag}$, significantly different than those reported in the SDSS archive, while the axial ratio and the position angle remain roughly constant in all bands with values of $\simeq0.56$ and $\simeq-46.5$, respectively.
Accounting for the redshift derived from the spectra of LSQ13zm and the foreground Galactic extinction reported by the NED archive, we infer an absolute magnitude $M_g=-15.99\pm0.21\,\rm{mag}$.
The low absolute luminosity already suggests a significantly sub-solar global metallicity of $12+\log{[\rm{O/H}]}=8.21\pm0.37$~dex, following the relation of \citet{2004A&A...423..427P}. \\

\begin{figure}
\includegraphics[width=1.0\columnwidth]{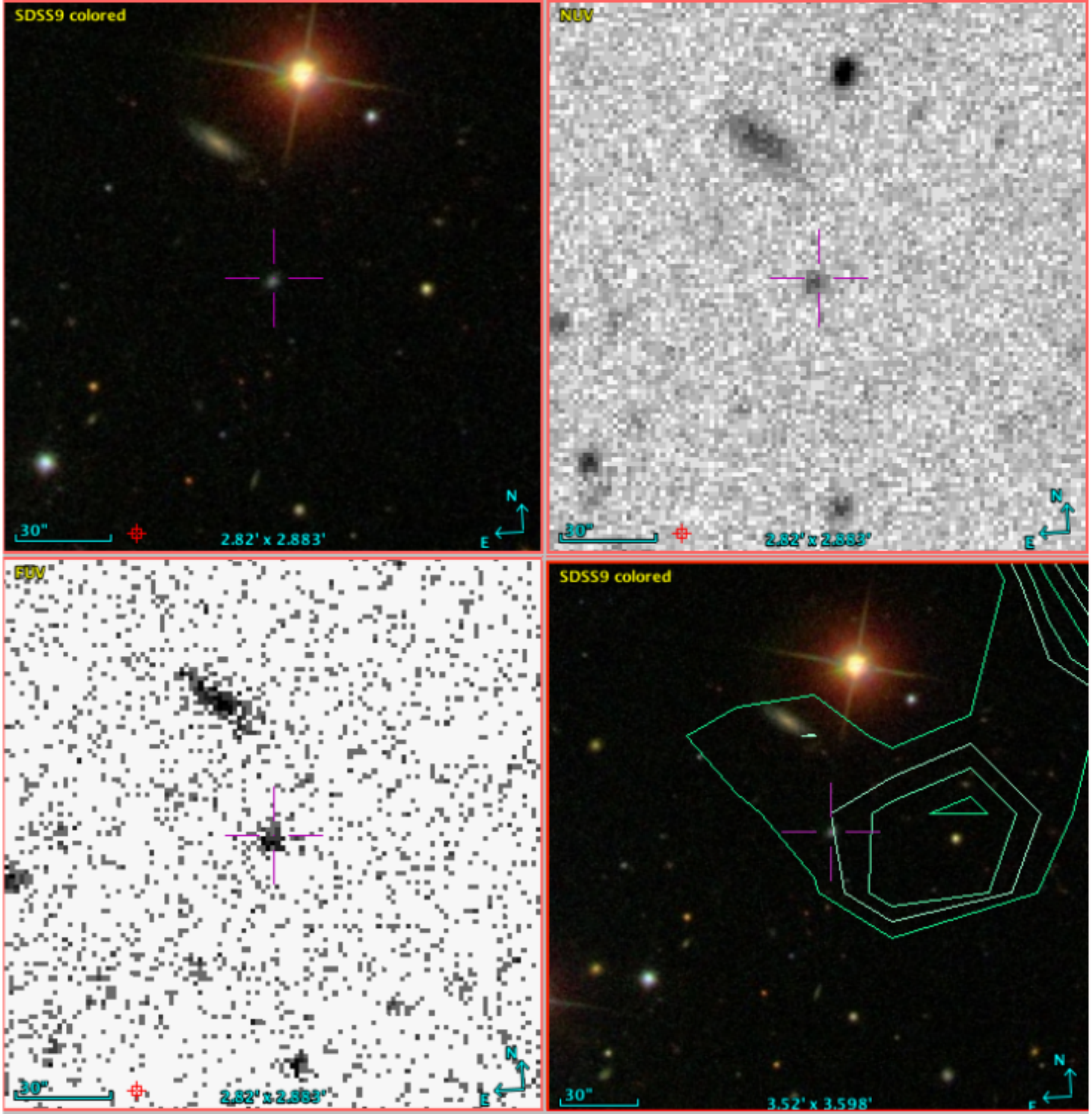}
\caption{Sky region of SDSS~J102654.56+195254.8 at different wavelength. {\bf Top, left:} Polychromatic SDSS image. {\bf Top, right:} NUV Galex frame. {\bf Bottom left:} Galex FUV image. {\bf Bottom right:} Polychromatic SDSS image with NVSS contour levels ($1.0\times10^{-3}$, $4.8\times10^{-4}$, $4.6\times10^{-4}$ and $4.3\times10^{-4}$~mJy respectively); in all the images the position of the galaxy is marked.
}
\label{galex}
\end{figure}

Global parameters were obtained using the \textsc{galfit\footnote{\url{http://users.obs.carnegiescience.edu/peng/work/galfit/galfit.html}}} code \citep{2010AJ....139.2097P}, hence performing 2D photometry fitting a Sersic law convolved with the local PSF.
The effective radii range from 1.2\arcsec~to 1.35\arcsec~in the different filters.
The results of this analysis confirmed our estimate on the actual position of LSQ13zm with respect to the centre of its host galaxy, that we assumed to be coincident with the centre of the isophotes (namely RA$=$10:26:54.638, Dec$=+$19:52:54.711~[J2000]).
Their nearly coincident positions, in particular, suggest that the progenitor star belonged to a stellar population located in the central regions of the host galaxy.
The Sersic index is low, ranging from 1.85 to 2.1 in the different bands.
This, together with the visual appearance, suggests that the host could be a very early spiral galaxy, an S0, or a low luminosity elliptical galaxy.
However, the SN spectra clearly show residual contamination of emission lines from a foreground \ion{H}{II} region (see Section~\ref{spectroscopy}). 
For this reason, we rule out the elliptical galaxy classification.
This choice is also supported by our estimated blue colours ($g-r=0.52$, $r-i=0.16$, $i-z=0.23$) as well as the clear detection of the host by the Galaxy Evolution Explorer (Galex\footnote{\url{http://www.galex.caltech.edu/}}), which also provides the relatively bright total magnitudes $21.49\pm0.14\,\rm{mag}$ and $21.64\pm0.12\,\rm{mag}$ in the Far--Ultraviolet (NUV, 1529~\AA) and Near--Ultraviolet (NUV, 2312~\AA) respectively.
In Figure~\ref{galex} we show multi-wavelength images of the field of SDSS~J102654.56$+$195254.8 obtained with SDSS, Galex and the NRAO VLA Sky Survey \citep[NVSS\footnote{\url{https://science.nrao.edu/science/surveys/vlass}};][]{1998AJ....115.1693C}. %%%
% Figure 3
Moreover, blue colours, along with \ion{[O}{II]}, \ion{[O}{III]} and \ion{[S}{II]} emission lines (clearly visible in the late time spectra of LSQ13zm, see Section~\ref{spectroscopy}) suggest high star formation rates (SFR), and are some of the common features of a class of galaxies known as `Blue Compact Dwarf Galaxies' (BCDGs).
BCDGs are active star--forming galaxies first identified by \citet{1956BOTT....2n...8H} and \citet{1961cgcg.book.....Z}.
They are characterised by blue colours and compactness, but also by low luminosities and metallicities, and usually host stellar super-clusters \citep{2011PhDT.......198A}.
Some of them show spectra similar to those of \ion{H}{II} regions of spiral galaxies, while the estimated SFRs range from 0.1 to 1~\msun~$\rm{yr}^{-1}$ \citep{1988ApJ...334..665F,2001ApJS..133..321C}.
According to this classification, and following \citet{2015MNRAS.451.2251I}, we estimated a global metallicity of $12+\log{[\rm{O/H}]}=7.93\pm0.003$~dex.
Low metallicities are common among the hosts of SN impostors, with values lower than those measured in galaxies hosting genuine Type IIn SNe \citep{2015arXiv150504719T,2010MNRAS.407.2660A}.
As a consequence, it is possible to argue that a fraction of interacting SNe does not belong to the same stellar population as SN impostors and that LBVs are not the only possible progenitor candidates for SNe IIn \citep{2015arXiv150504719T}. \\
The current SFR estimated from the FUV flux (9.2~$\mu$Jy) obtained using the \citet{1998ARA&A..36..189K} relation:
\begin{equation} \label{fuv_sfr}
  \rm{SFR}_{\rm{FUV}}~(\rm{M}_{\odot}~yr^{-1}) = 1.4\times10^{-28}~L_{\rm{FUV}}~(erg~s^{-1}~Hz^{-1}) 
\end{equation}
is 0.025~\msun~yr$^{-1}$, which is lower than the typical minimum value expected in this type of galaxies. 
We also computed an independent value using the integrated flux of the \ion{[O}{II]} 3727~\ang~host galaxy luminosity estimated from the $+24$~d LRS spectrum using the relation:
\begin{equation} \label{o2_sfr}
\rm{SFR}_\ion{[O}{II]}~(\rm{M}_{\odot}~yr^{-1}) = 1.4\times10^{-41}~L_{\ion{[O}{II]}}~(erg~s^{-1}) 
\end{equation}
derived by \citet{2002MNRAS.332..283R}, which gives a comparable low value of 0.028~\msun~yr$^{-1}$. 
Nonetheless, also the total mass derived adopting a `diet' Salpeter Initial Mass Function \citep[IMF][]{2001ApJ...550..212B} and using the $M/L$ -- colour relation given in \citet{2003ApJS..149..289B}, is relatively low, with a value (in stars) ranging from 8.65 to 8.8~Log($M/$\msun).
Adopting a \citet{2002Sci...295...82K} IMF, we derive an even lower mass ranging from 8.50 to 8.65~Log($M/$\msun), suggesting an high specific SFR. \\

The above described parameters show that SDSS~J102654.56+195254.8 is peculiar, since it shows compactness and morphological properties typically observed in early type galaxies, but also characterised by low metallicity and spectroscopic features typically observed in \ion{H}{II} regions of star-forming spiral galaxies.
The accurate characterisation of the host galaxies is an important tool in the study of peculiar transients, and could give an important improvement to our understanding of the physical processes occurring during the late phases of the evolution of massive stars.

%%%%%%%%
% SECTION 3
%%%%%%%%
\section{Photometry} \label{photometry}
The multi-band photometric monitoring campaign started on 2013 April 27, and spanned a period of above 200~d.
The 25 epochs of Sloan $g, r, i$ and $z$ photometry were primarily obtained using the 1 and 2-m telescopes of the Las Cumbres Observatory Global Telescope Network \citep[LCOGT\footnote{\url{http://lcogt.net/}};][]{2013PASP..125.1031B} and are reported in Table~\ref{grizCurves}. Near-Infrared (NIR) final magnitudes are reported in Table~\ref{JHKcurves}, while the observations in the Johnson--Cousins $B, V$ and $R$ bands were obtained using several facilities, all listed in Tables~\ref{BVRCurves} and \ref{photolimits}.
Historical limits and the first detections of the pre-SN outburst (the 2013a event), were obtained by the Catalina Sky Survey (CSS) 0.7~m Schmidt telescope, which provided up to $\sim10$~years of observations.
Additional data were provided by the LSQ and the iPTF surveys.
NIR data were obtained using the Rapid Eye Mount (REM) 0.6~m telescope with REMIR and the 2.54~m Nordic Optical Telescope (NOT) with NOTCam. 
The details about individual instrumental configurations are reported in the photometry tables (see Appendix). \\

\begin{figure} 
\includegraphics[width=1.0\columnwidth]{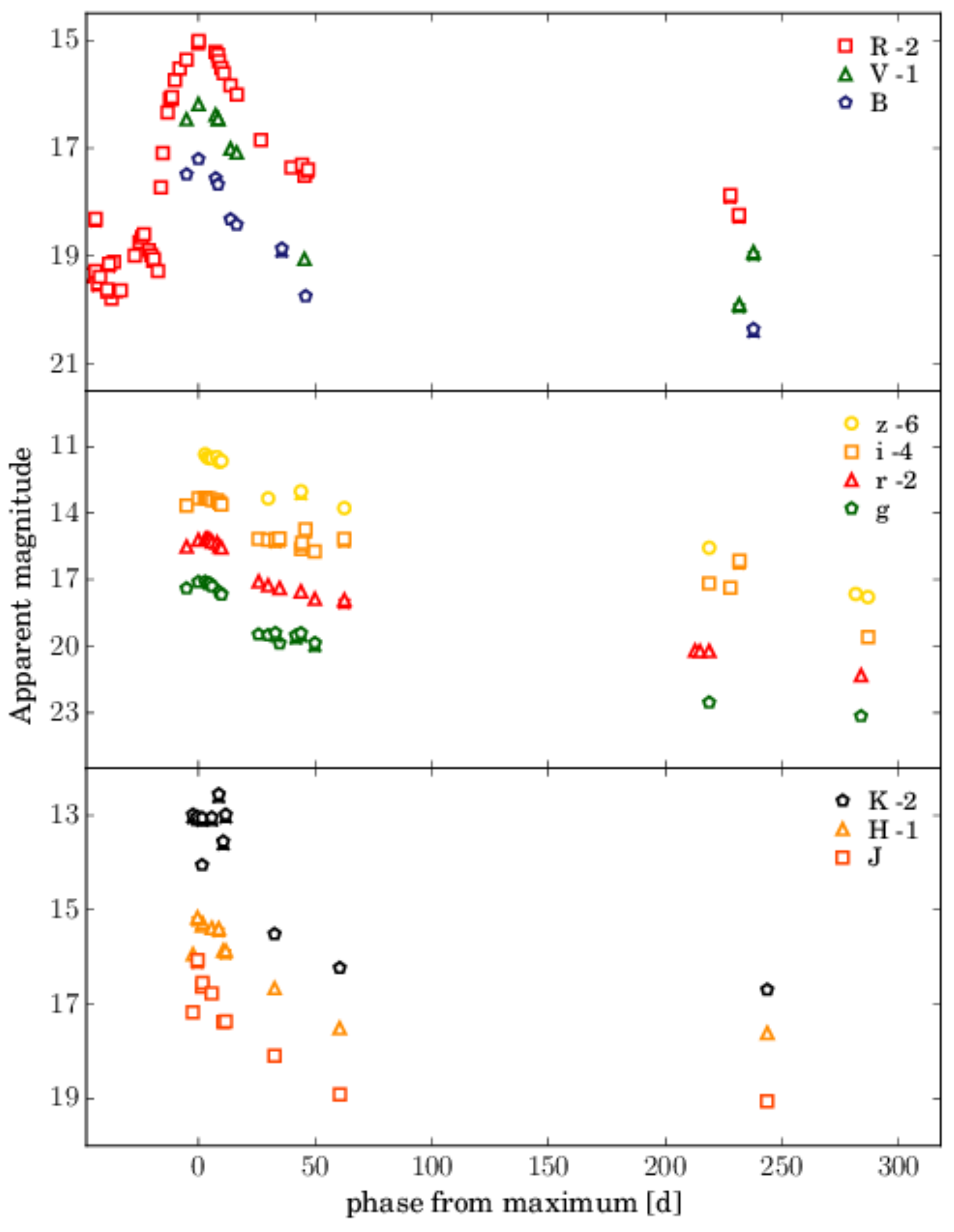}
\caption{Multi-band light-curves of the transient LSQ13zm. $BVR$ and $JHK$ magnitudes are calibrated to the Vega system, while $griz$ magnitudes to the AB photometric system. Arbitrary constants were applied to the magnitudes of different bands.}
\label{lightcurves}
\end{figure}

Photometric data were first pre-processed (applying overscan, bias and flat field corrections) using standard \textsc{iraf}\footnote{\url{http://iraf.noao.edu/}} procedures.
Multiple NIR exposures were optimised subtracting clear sky images obtained median-combining dithered images, and then the resulting images were combined to increase the signal--to--noise ratio (SNR).
The source instrumental magnitudes and their subsequent photometric calibration were obtained using a dedicated pipeline \citep[][SNoOpy]{snoopyref}. 
Template subtraction was performed with \textsc{hotpants} by PSF matching of the field stars, using archival SDSS images and very late images obtained with the GTC as templates.
No templates were available for NIR bands, hence instrumental magnitudes for these bands were obtained using the PSF-fitting technique on un-subtracted images.
Zero points and colour terms for the specific instrumental set-up were obtained with reference to a selected set of stars in the field.
The `local sequence' magnitudes were obtained from the SDSS and the 2MASS catalogues and were used to calibrate the SN magnitudes in the different images. 
BVR magnitudes of the reference stars were derived from the Sloan passband magnitudes, following \citet{2008AJ....135..264C}.
Photometric errors were estimated through artificial star experiments, combining in quadrature the dispersions of individual measurements with the PSF-fit errors returned by \textsc{daophot}.
For two epochs, multiband SNF synthetic photometry derived from flux-calibrated spectra was obtained using the procedures described in \citet{2013A&A...549A...8B}, while the specific LSQ and PTF data reduction procedures are described in detail in \citet{2015MNRAS.446.3895F}.
The final magnitudes of LSQ13zm along with the photometric errors are listed in Tables~\ref{grizCurves}, \ref{JHKcurves} and \ref{BVRCurves}, while the resulting light-curves are shown in Figure~\ref{lightcurves}.
% Figure 4
The first detection of the transient is dated 2013 March 18 ($\rm{MJD}=56369.109$) and was obtained by LSQ, followed by a marginal detection in a CRTS image obtained on 2013 March 19 ($\rm{MJD}=56370.150$).
After these epochs, the $R$-band magnitude rose to $\simeq20.6\,\rm{mag}$ (corresponding to $M_R\simeq-14.8\,\rm{mag}$, with the distance modulus and extinction values reported in Section~\ref{host}) until 2013 April 1 ($\simeq13$~d after the first detection), when the magnitude started to drop, reaching the value of $\simeq21.3\,\rm{mag}$ on 2013 April 7.
Unfortunately, no information of the colours was obtained for the 2013a event, since the target was followed in the $R$-band only and no spectra were collected at these epochs (Section~\ref{spectroscopy}).
On 2013 April 8 ($\rm{MJD}=56390.285$, the beginning of the 2013b event) a re-brightening was observed. 
After this epoch the $R$-band magnitude reached $\simeq17\,\rm{mag}$, corresponding to $M_R\simeq-18.4\,\rm{mag}$, on 2013 April 24 ($\rm{MJD}=56406.375$, the peak of the 2013b event), that we will consider hereafter (unless otherwise noted) as a reference for the phases in both the photometric and spectroscopic analysis.
During the re-brightening phase, we started to collect multi-band photometry and spectra using other facilities, as will be discussed in Sections~\ref{photoanalysis} and \ref{spectroscopy}.
Over the 20~d prior to the 2013b maximum, we measure a rise of $\sim2.5\,\rm{mag}$ in $R$-band, while after maximum the light-curves evolve faster with a $r$-band decline-rate of $r\sim5.6\,\rm{mag}$/100~d in the first $\sim25$~d, decreasing to $r\sim3\,\rm{mag}$/100~d until phase $\sim60$~d.
At later phases (namely after phase $+$200~d) we notice a slower decline, with a rate of $r\sim0.2\,\rm{mag}$$/100$~d in the $r$-band light-curve, while the slopes in the $g$,$i$ and $z$-bands are greater: 0.96, 3.62 and 3.28~mag$/100$~d respectively. 
This can be explained with an increased contribution of the \halpha~emission line relative to the continuum, although it has to be mentioned that we have no observations between phases $+50$ and $+219$~d in $g$ and $i$-band, and between phases $+63$ and $+219$~d in $z$-band.

%%%%%%%%%
% SECTION 3.1
%%%%%%%%%
\subsection{Absolute light-curves and colour curves} \label{photoanalysis}
\begin{figure} 
\includegraphics[width=1.0\columnwidth]{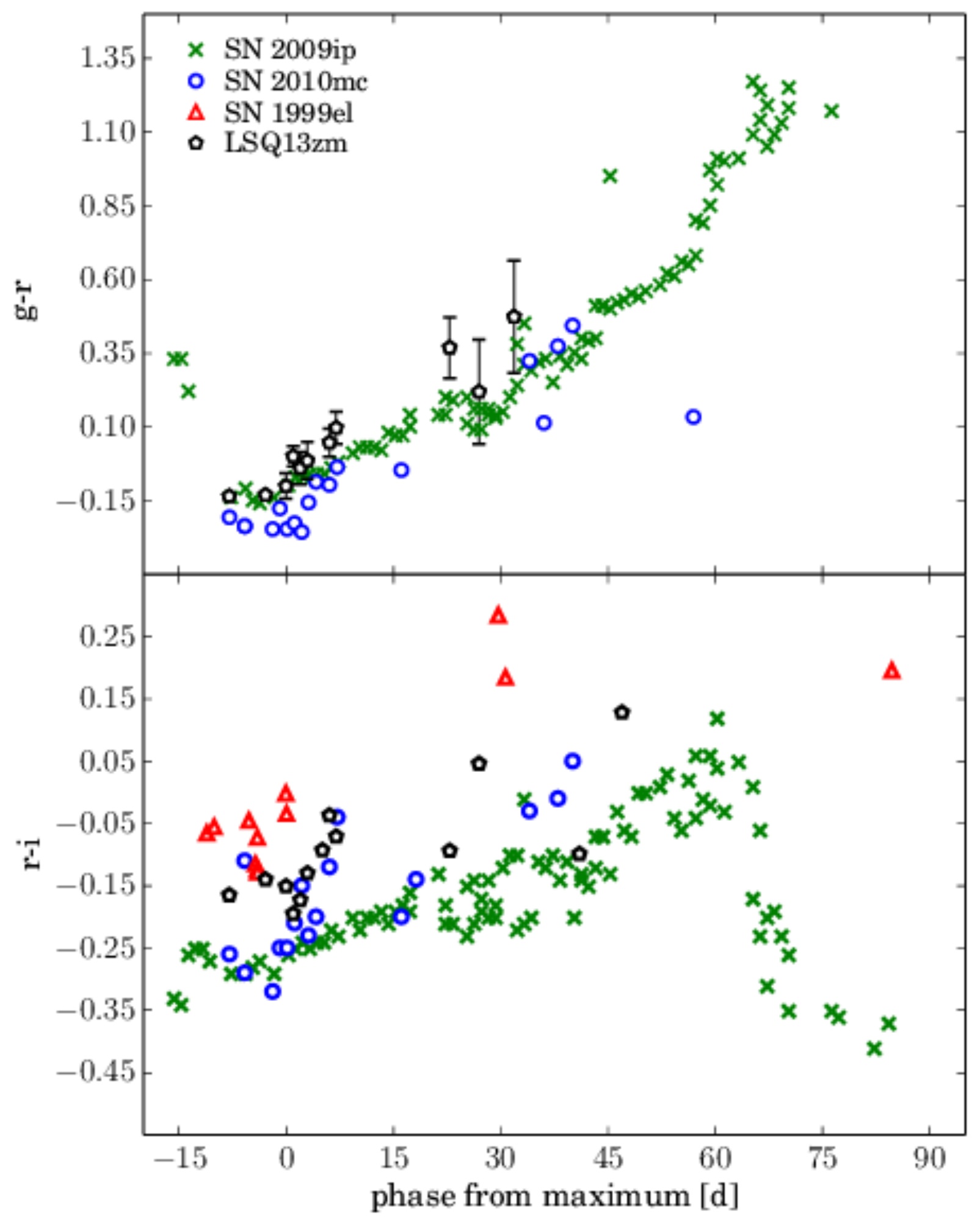}
\caption{Comparison among the $g-r$ and $r-i$ colour curves of LSQ13zm, SN~2010mc, SN~1999el and SN~2009ip. Reddening estimates of SN~2009ip ($A_V=0.055\,\rm{mag}$) and SN~2010mc ($A_V=0.046\,\rm{mag}$) were obtained from the NED archive. For the total extinction to the direction of SN~1999el we adopted the value $A_V=1.84\,\rm{mag}$, an average of the two extreme values reported by \citet{2002ApJ...573..144D}. Magnitudes were calibrated on the AB photometric system. The phases of SN~2009ip are relative to the 2012b event.}
\label{colCurves}
\end{figure}

Figure~\ref{colCurves} shows the $g-r$ (top) and $r-i$ (bottom) colour evolution of LSQ13zm along with those of SN~2010mc \citep{2013Natur.494...65O} and SN~2009ip, two objects showing a similar photometric behavior, and the Type IIn SN~1999el \citep{2002ApJ...573..144D}.
% Figure 5
The $g-r$ and $r-i$ colours become progressively redder with time, suggesting a rapid temperature decrease of the ejecta, as will be shown in the spectral analysis (Section~\ref{spectroscopy}). \\

\begin{figure} 
\includegraphics[width=1.0\columnwidth]{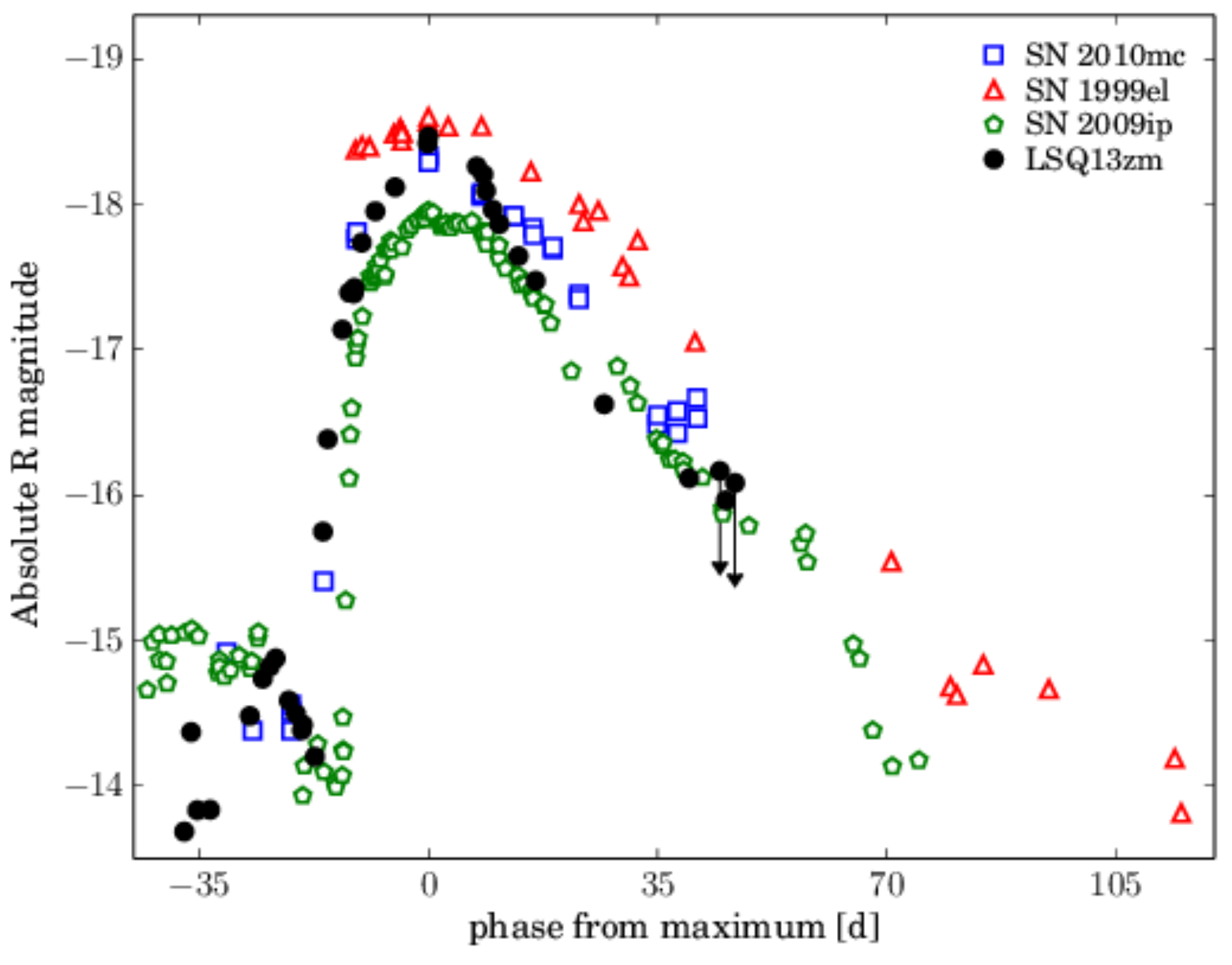}
\caption{Comparison among the $R$-band absolute light curves of SN~1999el, SN~2010mc, SN~2009ip and LSQ13zm. The phases are relative to the maximum of the brightest event. The distance moduli of SN~1999el ($\mu=32.1\,\rm{mag}$), SN~2009ip ($\mu=31.55\,\rm{mag}$) and SN~2010mc ($\mu=35.79\,\rm{mag}$) were taken from \citet{2002ApJ...573..144D,2010AJ....139.1451S} and \citet{2013Natur.494...65O}, respectively.}
\label{absComparison}
\end{figure}

The absolute $R$-band light-curve of LSQ13zm, with phases relative to the 2013b event, is compared with those of the same SN sample as above in Figure~\ref{absComparison}. 
Adopting for LSQ13zm the distance modulus and extinction discussed in Section~\ref{host}, we estimate an absolute peak magnitude of $M_R=-18.46\pm0.21\,\rm{mag}$ for the 2013b event.
% Figure 6
As shown by this comparison, the absolute light-curve of the 2013b event of LSQ13zm is reminiscent of those of other Type IIn SNe, showing similar absolute peak magnitudes \citep[in agreement with][who found $-18.4\,\rm{mag}$ as a mean value for the peak magnitudes for SNe IIn]{2012ApJ...744...10K} and decline rates.
In particular, there is a remarkable similarity between the 2013b absolute light-curve of LSQ13zm and that of SN~2010mc. \\

\begin{figure} 
\includegraphics[width=1.0\columnwidth]{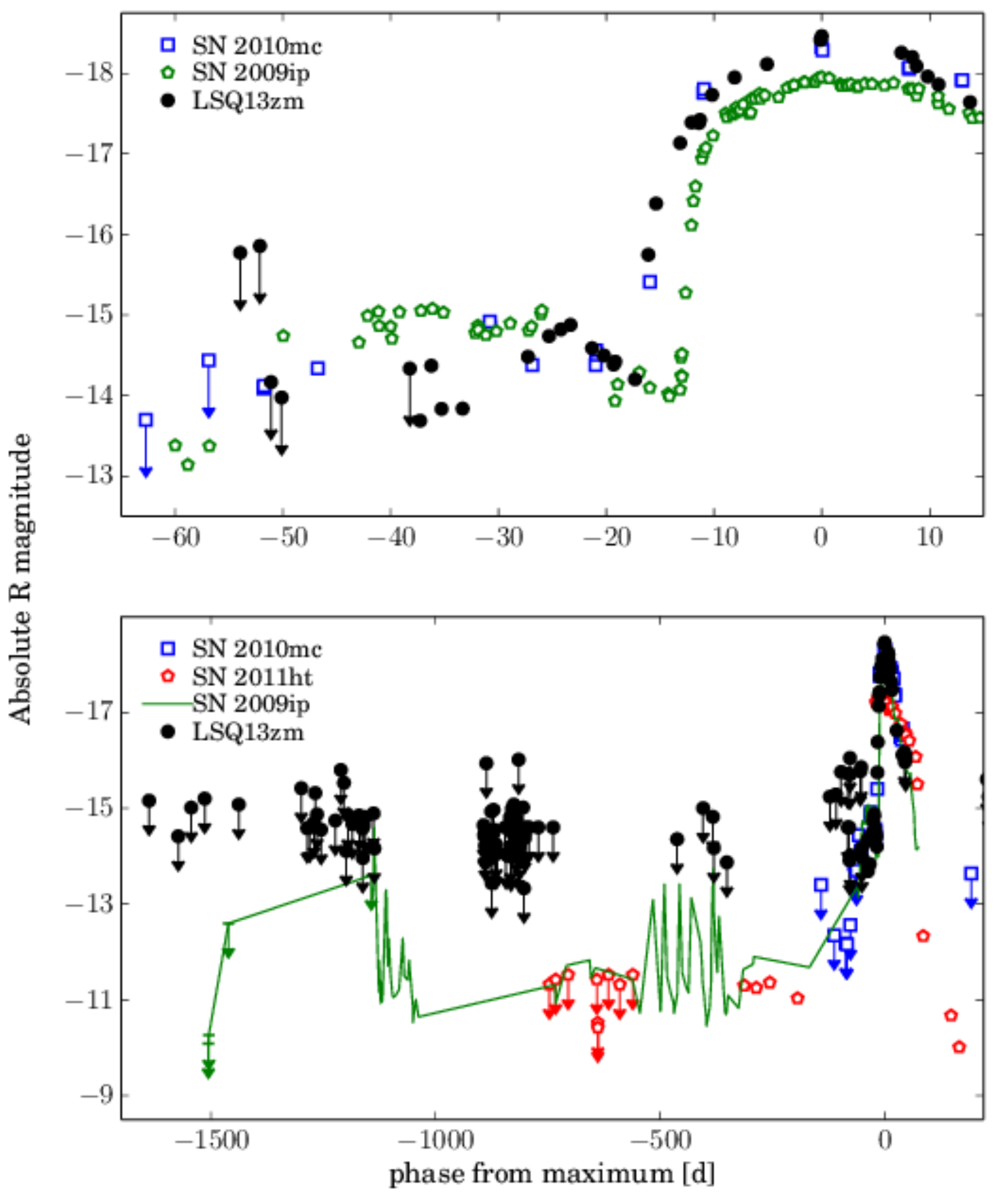}
\caption{Comparison of the long-term $R$-band absolute light curves of LSQ13zm, SN~2011ht and SN~2009ip. {\bf Top:} blow-up of the pre-SN outbursts of LSQ13zm (2013a) and SN~2009ip \citep[2012a,][]{2013ApJ...767....1P}. {\bf Bottom:} comparison of historical observations of LSQ13zm, SN~2011ht and SN~2009ip. References are indicated in the main text. The indicated phases are relative to the maximum of the most luminous event, possibly corresponding to the final SN explosion. The distance modulus of SN~2011ht ($\mu=31.42\,\rm{mag}$) was taken from \citet{2012ApJ...751...92R} while, the reddening estimate ($A_V=0.029\,\rm{mag}$) is from the NED archive.
}
\label{abslc}
\end{figure}

In Figure~\ref{abslc}, we show the long-term photometric evolution of LSQ13zm (including the pre-discovery phases). 
Its $R$-band absolute light curve is compared with those of other objects classified as Type IIn SNe showing pre-explosion outbursts.
Adopting the same values for the distance modulus and extinction, we infer an absolute peak magnitude of $M_R=-14.87\pm0.25\,\rm{mag}$ for the 2013a event of LSQ13zm, comparable with the 2012a event of SN~2009ip. \\

Figure~\ref{abslc} also reports observations of the LSQ13zm site obtained prior to the 2013a,b episodes, including the photometric detection limits collected from CTRS, LSQ and iPTF archival images. 
The top panel shows that the pre-SN bursts of LSQ13zm, SN~2009ip and SN~2010mc are quite similar, with comparable absolute peak magnitudes, although the 2013a episode of LSQ13zm has a shorter duration.
The Type IIn SN~2011ht (bottom panel) is slightly different, showing a pre-SN burst occurred $\sim1$~year before the SN explosion \citep{2013ApJ...779L...8F}.
The bottom panel shows that no transient was observed in the past decade at the position of LSQ13zm, although the detection limits constrain the non-detections only to absolute magnitudes in the range between $M_R\approx-13.5\,\rm{mag}$ and $-15\,\rm{mag}$.
Of course, these observations cannot rule out that previous outburst episodes occurred in the gaps between the observations or at a fainter magnitude. 
In fact the detection limits of LSQ13zm (prior to the 2013a event) are not very stringent, as they are typically brighter than the erratic bursts observed in SN~2009ip during the period 2009-2011 \citep{2013ApJ...767....1P}. \\
% Figure 7

%%%%%%%$
% SECTION 4
%%%%%%%%
\section{Spectroscopy} \label{spectroscopy}
Our spectroscopic follow-up campaign started on 2013 April 19 and lasted until 2014 November 23, hence covering almost 2~years.
Essential information about the spectra is reported in Table~\ref{speclog}. 
The spectra will be released through the Weizmann Interactive Supernova data REPository \citep[WISeREP\footnote{\url{http://wiserep.weizmann.ac.il/}};][]{2012PASP..124..668Y}. \\
% Table 1
\begin{table*}
\begin{minipage}{175mm}
\caption{Log of the spectroscopical observations.}
\label{speclog}
\begin{tabular}{@{}cccccccc@{}}
\hline
Date & MJD & Phase & Instrumental setup & Grism or grating & Spectral range & Resolution & Exp. times   \\ 
        &          & (d) &                               &                            & (\ang)                  & (\ang)           & (s)                \\ 
\hline
20130419 & 56401.3 & $-5$ &      2.2~m UH+SNIFS  &  Channels B$+$R & 3180-9420 &   9.7 &  $2\times1820$ \\
20130425 & 56406.4 & 0 &      2.2~m UH+SNIFS   & Channels B$+$R & 3180-9420 &   9.4 &  $2\times1820$ \\
20130428 & 56410.3 & $+4$ &     FLOYDS                   & Channels red$+$blue & 3200-6600 & 8.9 & $2\times3600$ \\
20130502 & 56414.5 & $+8$ &      BTA6~m+SCORPIO  &  VPHG550G  &  3100-7300 &  12.7 &  2550 \\
20130508 & 56421.3 & $+15$ &      Ekar182+AFOSC  &  Gm4 &  3600-7900 &  24 &  2700 \\
20130510 & 56422.5 & $+16$ &      BTA6~m+SCORPIO  &  VPHG550G & 3600-7660 &  12.7 &  3600 \\
20150517 & 56429.9 & $+24$ &      TNG+LRS &       LR-B &  3200-7700 &  11.9 &  3600 \\
20130523 & 56435.9 & $+30$ &      TNG+LRS &       LR-B & 3200-7700 &  11.1 &  3000 \\
20130606 & 56449.9 & $+44$ &      TNG+LRS &       LR-B &  3200-7700 &  15.2 &  3000 \\
20130629 & 56472.9 & $+67$ &      TNG+LRS &       LR-B &  3200-7700 &  18.8 &  $2\times2700$ \\
20131225 & 56651.9 & $+245$ &      GTC+OSIRIS &       R500R &  4670-9000 &  16.2 &  1800 \\
20141124 & 56985.2 & $+579$ &      GTC+OSIRIS &       R300B &  4000-9770 &  17.1 &  $2\times1800$ \\
\hline
\end{tabular}

\medskip
The observations were carried out using the 6.05~m Bolshoi Teleskop Alt-azimutalnyi (BTA) of the Special Astronomical Observatory equipped with the Spectral Camera with Optical Reducer for Photometrical and Interferometrical Observations (SCORPIO; Zelenchuksky District, Caucasus Mountains, Russia), 2.2~m telescope of the University of Hawaii with the SuperNova Integral Field Spectrograph (SNIFS; Mauna Kea, Hawaii), the 10.4~m Gran Telescopio Canarias (GTC) with OSIRIS and the 3.58~m Telescopio Nazionale Galileo (TNG) with DOLoRes (LRS; both located at Roque de Los Muchachos, La Palma, Canary Islands, Spain), the 1.82~m Copernico telescope with AFOSC (Mt.~Ekar, Asiago, Italy) and the LCOGT 2~m Faulkes North Telescope with FLOYDS.
\end{minipage}
\end{table*}

One-dimensional spectra were obtained using standard \textsc{iraf} tasks for both pre-reduction (bias, flat field and overscan correction) and optimised extraction.
Wavelength calibration was performed using the spectra of comparison lamps obtained during the same night. 
The accuracy of the wavelength calibration was then verified measuring the positions of night sky lines, in particular \ion{[O}{I]} at 5577.34~\ang~and 6300.30~\ang, and shifting the spectrum in wavelength in case of discrepancy.
Flux calibration was performed using spectra of standard stars. 
Measured fluxes were checked against multi-band photometry obtained on the nearest nights and, when necessary, a scaling factor was applied. 
We did not perform any flux correction on the last two OSIRIS spectra, since they were both strongly contaminated by the flux of the host galaxy.
Spectral resolutions, reported in Table~\ref{speclog}, were computed measuring the mean values of the full--widths--at--half--maximum (FWHM) of unblended night sky lines.

%%%%%%%%%
% SECTION 4.1
%%%%%%%%%
\subsection{Line identification and spectral evolution} \label{linesIDspecEV}
\begin{figure*} 
\includegraphics[width=0.65\linewidth]{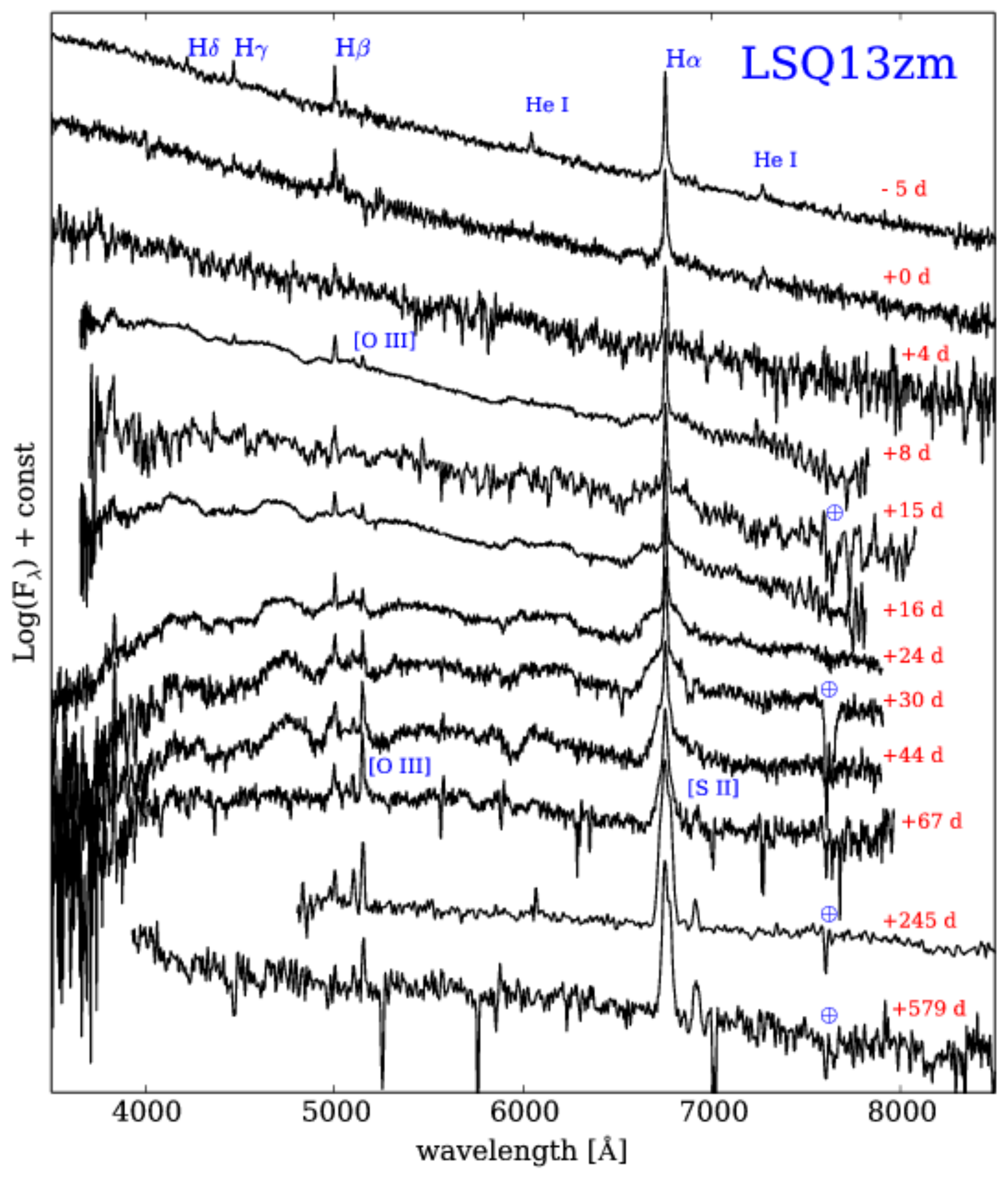}
\caption{Spectral sequence of LSQ13zm. The phases, in days, are reported to the right, and are relative to the light-curve maximum of the 2013b event. The $\oplus$ symbols mark the positions, where visible, of the strongest telluric absorption bands. All spectra are flux-calibrated using the information obtained from the photometric data, with the exception of the last two OSIRIS spectra for which the galaxy contamination could not be removed. The spectra have been shifted by an arbitrary constant. The strongest lines visible in emission are also marked. The spectra are in the observed frame, and no reddening correction has been applied.}
\label{specSeq}
\end{figure*}

The full spectral sequence of LSQ13zm is shown in Figure~\ref{specSeq}.
% Figure 8
The spectra show a blue continuum that becomes progressively redder with time, as the temperature at the photosphere decreases. 
A significant evolution in the relative strengths of the spectral features can also be noted.
The first two spectra are dominated by narrow Balmer lines in emission, probably arising from a photo-ionised thin shell expelled a relatively short time ahead of the 2013b event.
Around $+8$~d, the spectrum begins to show broad shallow features with P-Cygni profiles, most clearly visible in \halpha~and \hbeta.
From $+16$~d onwards, broad P-Cygni absorptions dominate the spectrum, including those of \hbeta~and other blended lines, as will be discussed in the following sections.
We also identify blended \ion{Fe}{II}, \ion{Mg}{II} and \ion{Ca}{II} lines, taking as a comparison the super-luminous Type Ic SN~2010gx \citep[Figure~\ref{cfr_spec};][]{2010ApJ...724L..16P}.
From $+44$~d to $+67$~d, the interaction of the ejecta with the CSM becomes dominant, resulting in a progressively stronger contribution of the \halpha~intermediate component, which merges with the broad line wings of the profile. 
At $+67$~d, the intermediate component dominates the \halpha~profile over the other components. \\

In the early spectra, we marked the two \ion{He}{I} lines in emission at 5875.6 and 7065.2~\ang~(see Figure~\ref{specSeq}), which are clearly detectable until $+8$~d. 
On the basis of the available data (resolution and SNR), we cannot rule out the presence of a weak \ion{He}{I} (5875.6~\ang) line in absorption also at later phases. 
At $+8$~d, we note the presence of the \ion{[O}{III]} $\lambda\lambda$4958.5, 5006.8 doublet, although the single lines remain un-resolved. 
In addition, at late phases, also the \ion{[S}{II]} $\lambda\lambda$6717-6731 doublet becomes detectable. 
Unresolved \ion{[S}{II]} lines are uncommon in the nebular spectra of CCSNe, suggesting that they are an evidence of host galaxy contamination. 
Although these unresolved forbidden lines appear to increase in flux with time, this is likely a consequence of the growing contamination of the galaxy foreground sources. 
This is clearly shown in Figure~\ref{spec_2d}, where the latest two-dimensional spectrum of LSQ13zm including the host galaxy is shown for two wavelength windows: the \halpha~+~\ion{[S}{II]} 6717-6731~\ang~doublet (top panel in the figure), and \hbeta~+~\ion{[O}{III]} $\lambda\lambda$4958.5, 5006.8 doublet (bottom panel).
In both cases, while the H lines are broader, \ion{[O}{III]} and \ion{[S}{II]} are narrower and spatially offset with respect to the peak of the H emissions.
The relatively strong and unresolved \ion{[O}{II]} $\lambda\lambda$3726-3729 doublet is also visible in the spectra from $+24$~d to $+67$~d, most likely due to the host contamination as well. \\

\begin{figure}
\includegraphics[width=1.0\columnwidth]{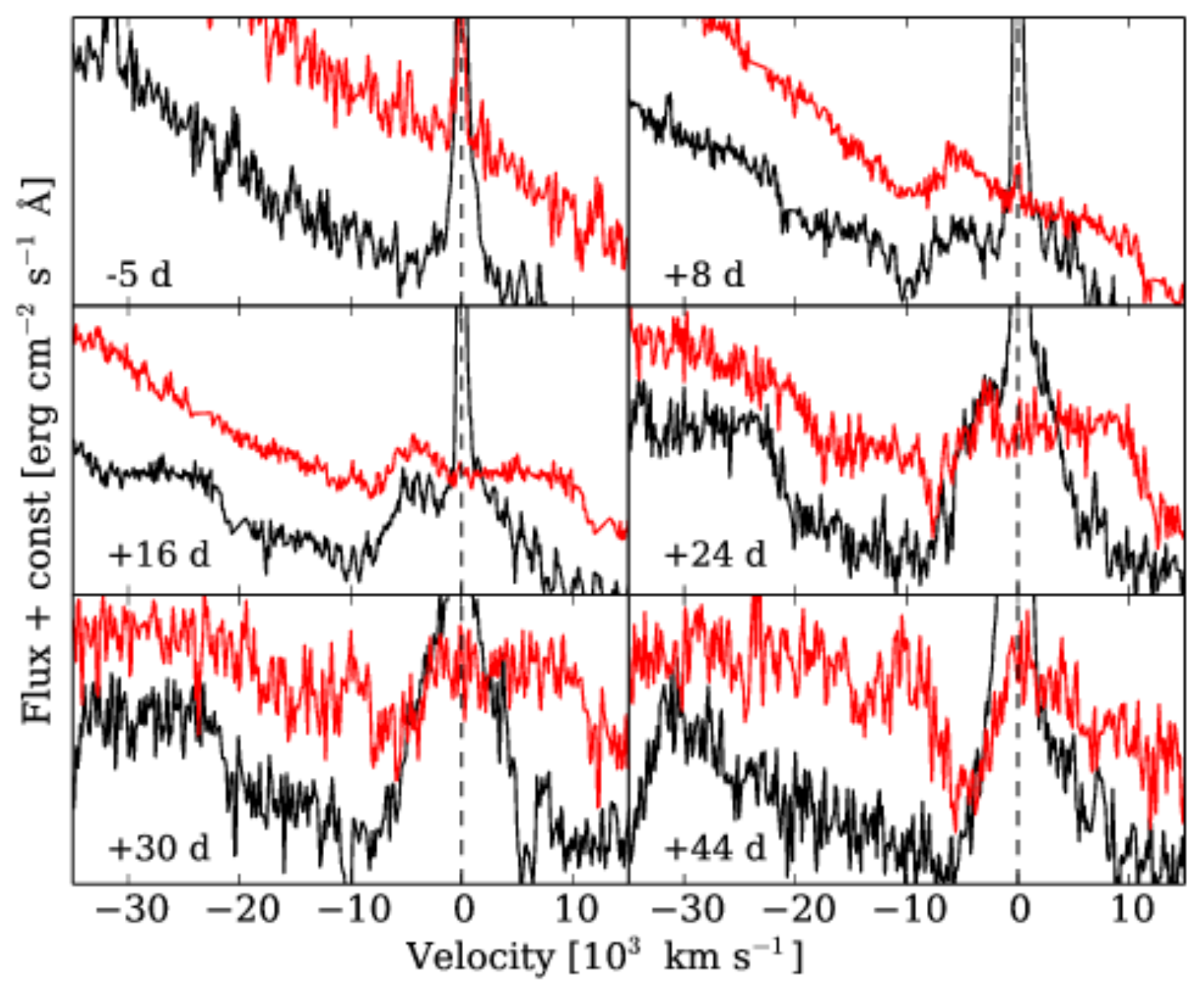}
\caption{Blow-up of the H$_{\alpha}$ (black line) and \ion{He}{I}/\ion{Na}{ID} features (red line) at $=-5,+8,16,24,30$ and 44~d, respectively. The dashed line is centred at the corresponding zero velocity of the \halpha~and \ion{He}{I} 5875.6\ang~transitions. The fluxes are re-scaled and a constant is applied.}
\label{hhedetail}
\end{figure}

From $+8$~d to $+67$~d, we identify a very broad \halpha~component and, at the location of the \ion{Na}{ID} doublet, another shallow feature in absorption showing a `boxy' profile. 
The close-by comparisons in Figure~\ref{hhedetail} show the temporal evolution of these two spectral regions in the velocity space, where the absorption features observed in \halpha~and \ion{He}{I}/\ion{Na}{ID} are compared. We note that broad P-Cygni absorptions are clearly visible in the \halpha~profile since phase $+8$~d, indicating the presence of a significant amount of material moving at $\simeq10^4$\kms, with a maximum inferred velocity of $\simeq22000$\kms.
The existence of fast-moving material is also confirmed by the absorption features observed in the \ion{He}{I}/\ion{Na}{ID} region.
The range of velocities inferred from the two different lines at $+44$~d also agrees, with a velocity $\simeq5\times10^3$\kms~for the bulk of the ejected material, and a maximum velocity of the line wing of $\simeq10^4$\kms, indicating that they arise from the same regions.
We also notice a second absorption component in the \ion{He}{I} (5876~\ang)/\ion{Na}{ID} (5891, 5897~\ang) P-Cygni profile.
This could be explained with the blended contribution of the two \ion{He}{I} and \ion{Na}{ID} lines.
Such high velocity absorption features are generally observed in the ejecta of CCSNe at early phases, and are not typical of LBV eruptions. 
This is one of the main arguments that supports the SN scenario for the 2013b event (see Section~\ref{discussion}). \\
% Figure 9

\begin{figure} 
\includegraphics[width=1.0\columnwidth]{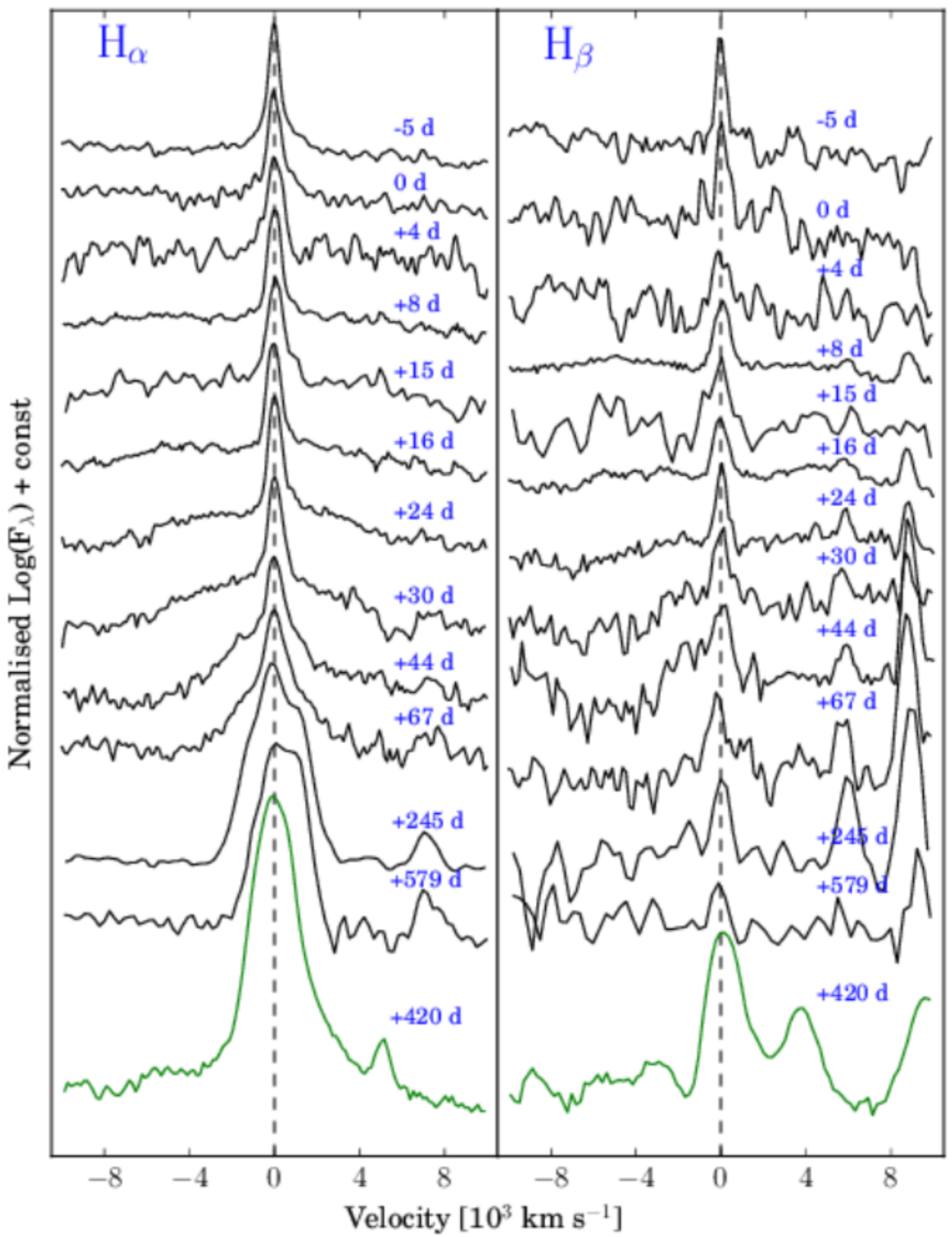}
\caption{Evolution of the profiles of \halpha~({\bf left panel}) and \hbeta~({\bf right panel}) in the velocity space. The fluxes are re-scaled and a constant is applied. The vertical dashed lines mark the rest wavelength positions of the two lines. A late-phase spectrum of SN~2009ip is also shown for comparison \citep{2015MNRAS.453.3886F}. The fluxes of the two lines were normalised and then divided by a factor 2 (\halpha) and 4 (\hbeta) in the SN~2009ip spectrum to facilitate the comparison.}
\label{hprofiles}
\end{figure}

Figure~\ref{hprofiles} displays the evolution of the \halpha~and \hbeta~emission line profiles.
% Figure 10
These two features are the only Balmer lines visible in all the spectra of our sequence, since \hgamma~disappears after a few days.
We do not observe a significant shift in the position of the broad \halpha~and \hbeta~components, very likely arguing against a prompt dust formation.
A spectrum of SN~2009ip taken at very late phases is also shown as a comparison. \\

\begin{figure}
\includegraphics[width=1.0\columnwidth]{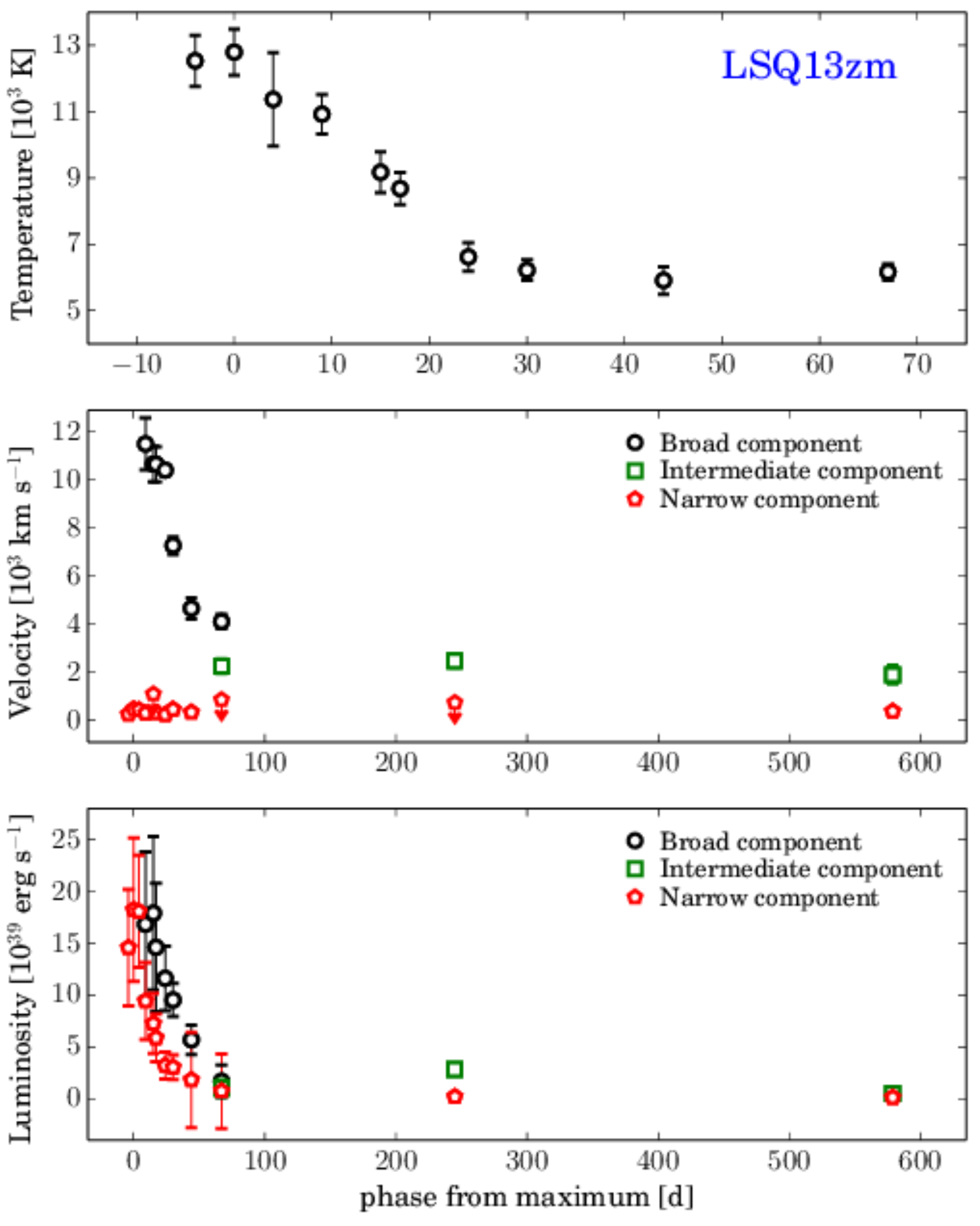}
\caption{{\bf Top:} Evolution of the temperature of the ejecta obtained through black-body fits. {\bf Middle:} FWHM evolution for the broad (black dots), intermediate (green squares) and narrow (red pentagons) of the \halpha~components. {\bf Bottom:} Luminosity evolution of the three \halpha~components.}
\label{tempfluxspeed}
\end{figure}

Figure~\ref{tempfluxspeed} shows the evolution of a few spectral parameters, viz. continuum temperature (top), velocity (middle) and luminosity (bottom) of the three line components, after correcting the spectra for foreground extinction and redshift.
Their values are listed in Table~\ref{physTab}.
% Figure 11
The velocities for the different line components are obtained through a multicomponent fit, using Gaussian and Lorentzian functions to reproduce the different \halpha~emission profiles.
In the two earliest spectra, we obtain a good fit is reached using a single Lorentzian component, while in all other spectra a fair result is obtained using a combination of multiple Gaussians.
This choice is based on the fit results, mostly affected by different resolution and SNR of the spectra.
When the different components have widths which are larger than the resolution of the spectra, we compute the final velocity $v=\frac{width}{\lambda_0}\times c$ from the measured width corrected for the spectral resolution ($width=\sqrt{\rm{FWHM}^2-\rm{res}^2}$).
When the components are un-resolved, we adopt the resolution of the spectra as an upper limit of the velocities.
For the narrow component we infer a nearly constant velocity, ranging from $250\pm10$\kms~to $470\pm20$\kms~(but strongly affected by spectral resolution limitations), while the broad component has a fast decline in velocity at early phases, from $11500\pm1080$\kms~to $4110\pm300$\kms~in the first 67~d. \\

From $+67$~d onwards, we include an intermediate component with velocities declining from $2560\pm350$\kms~to $1900\pm400$\kms.
As discussed in \citet{1993MNRAS.265..471T} and \citet{2012ApJ...744...10K}, spectra with multicomponent line profiles are common in interacting objects.
The narrow line components have velocities consistent with those expected for the material expelled from an LBV, and are interpreted as recombination lines emitted by an outer un-shocked CSM.
The FWHM velocities of the broad component are consistent with those measured in the expanding ejecta of a CCSN, although occasionally large velocities have been observed during major eruptions of LBVs \citep[see e.g. SN~2009ip,][during the period 2009-2011]{2013ApJ...767....1P}.
Finally, the velocities of the intermediate component are fully consistent with those predicted for shocked gas, lying in the region between the forward and the reverse shock fronts. \\

The evolution of the temperature of the photosphere (Figure~\ref{tempfluxspeed}, top) is inferred through a black-body fit to the spectral continuum, since the spectra at early and intermediate phases (from $+30$~d to $+67$~d) are all characterised by a significant contribution of the continuum to the total emission.
The temperature rapidly decreases from $12540\pm770$~K to $6170\pm250$~K in the first two months after maximum light.
We do not compute a temperature in the last two spectra, since at this phase the continuum is strongly contaminated by the emission of the host galaxy.
The temperature evolution is consistent with that of the broad-band colours discussed in Section~\ref{photoanalysis} (see also Figure~\ref{colCurves}). \\

We also derive the luminosity evolution for the different components of \halpha~(Figure~\ref{tempfluxspeed}, bottom).
The luminosity declines from $1.50\times10^{40}$ to $1.30\times10^{38}$~\ergs~for the narrow component, from $1.80\times10^{40}$ to $1.80\times10^{39}$~\ergs~for the broad component and from $1.10\times10^{39}$ to $5.50\times10^{38}$~\ergs~for the intermediate components over the period of the spectroscopic monitoring. \\

% Table 2
\begin{table*}
\begin{minipage}{175mm}
\caption{Main spectral parameters, as result from the analysis performed on the H$_{\alpha}$ line profiles.}
\label{physTab}
\begin{tabular}{@{}cccccccc@{}}
\hline
phase & Temperature & FWHM$_{\rm{nar}}$ & FWHM$_{\rm{brd}}$ & FWHM$_{\rm{int}}$ & L$_{\rm{nar}}$ & L$_{\rm{brd}}$ & L$_{\rm{int}}$ \\
  (d)     & (K)                & (\kms)                       & (\kms)                            & (\kms)                     & (\ergs)                              & (\ergs)                                 & (\ergs)                           \\
\hline
$-5$      &12540(770)   & $\sim250$*  & --                   & --              & $1.50(0.55)\times10^{40}$ & --                                          & -- \\
0           &12800(700)   & 470*(20)  & --                   & --              & $1.90(0.70)\times10^{40}$ & --                                          & -- \\
$+4$     &11370(1410) & 440(50)  & --                   & --              & $1.90(0.50)\times10^{40}$ & --                                          & -- \\
$+8$     &10930(590)   & 300(10)  & 11500(1080) & --              & $9.90(0.40)\times10^{39}$ & $1.80(0.30)\times10^{40}$  & -- \\
$+15$   & 91780610)    & $<1100$ & 10670(740)  & --              & $7.60(0.30)\times10^{39}$ & $1.90(0.30)\times10^{40}$  & -- \\
$+16$   & 8680(490)    & $\sim$330  & 10660(730)  & --              & $6.20(0.35)\times10^{39}$  & $1.50(0.20)\times10^{40}$  & -- \\
$+24$   & 6620(430)    & $\sim250$    & 10410(720)  & --              & $3.40(0.30)\times10^{39}$  & $1.20(0.30)\times10^{40}$  & -- \\
$+30$   & 6230(310)    & 460(10)  &  7280(440)   & --              & $3.20(0.40)\times10^{39}$  & $1.00(0.40)\times10^{40}$  & -- \\
$+44$   & 5920(410)    & $\sim340$  &  4660(560)   & --              & $1.90(0.50)\times10^{39}$  & $6.00(0.35)\times10^{39}$  & --   \\
$+67$   & 6170(250)   & $<8600$ &  4110(300)    & 2560(350) & $7.70(0.40)\times10^{38}$ & $1.80(0.30)\times10^{39}$ & $1.10(0.50)\times10^{39}$ \\
$+245$ & --                 & $<740$   &   --                 & 2490(280) & $2.00(0.30)\times10^{38}$ & --                                         & $3.00(0.60)\times10^{39}$ \\
$+579$ & --                 & 380(10)   &  --                  & 1900(400) & $1.30(0.60)\times10^{38}$  & --                                       & $5.50(0.50)\times10^{39}$ \\
\hline
\end{tabular}

\medskip
Column 1 reports the phases relative to the light-curve maximum, column 2 lists the temperatures derived through a black-body fit on the spectra continuum, columns 3,4 and 5 report the FWHM velocities inferred from multi-component fits of the \halpha~emission profiles for the narrow (nar), broad (brd) and intermediate (int) components, respectively. Columns 6,7 and 8 report the total luminosities inferred from the same \halpha~emission line components. We remark that values for the luminosities of the narrow and intermediate components in the last two spectra (phases $+245$ and $+579$~d are strongly affected by the contamination of the host galaxy emission lines. The * symbol refers to the velocities inferred through a Lorentzian fit only. The $\sim$ symbol refers to narrow lines marginally resolved.
\end{minipage}
\end{table*}

%%%%%%%%%
% SECTION 4.2
%%%%%%%%%
\subsection{Comparison with spectra of interacting SNe} \label{spec_compare}
\begin{figure} 
\includegraphics[width=1.0\columnwidth]{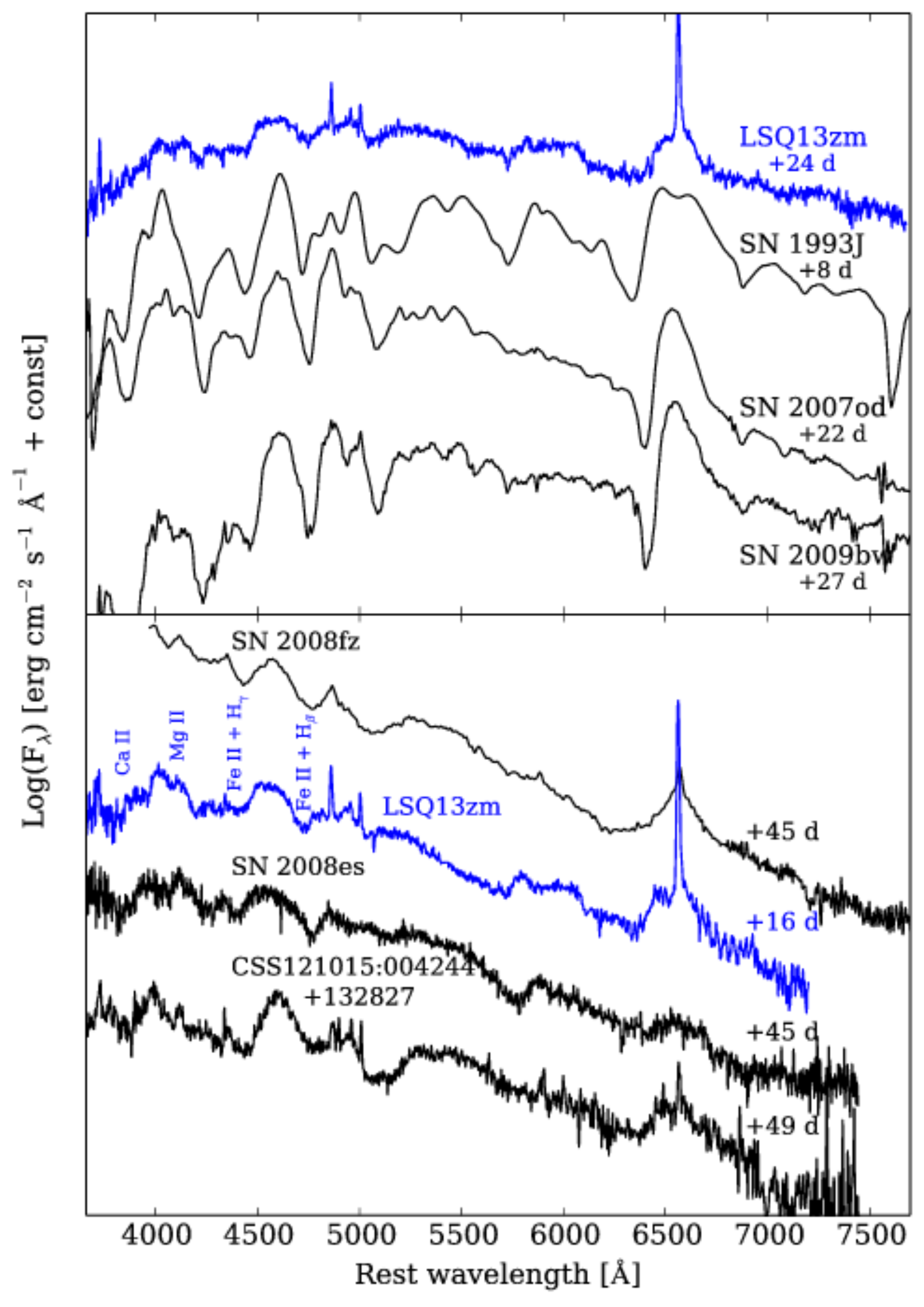}
\caption{Comparison of the $+16$~d and the $+24$~d spectra with those of a sample of CCSNe and SLSNe. The objects were selected on the basis of good fits obtained using \textsc{gelato}. Different constants have been applied to the logarithm of the fluxes.}
\label{cfr_spec}
\end{figure}

A comparison with the spectra of different CCSNe is shown in Figure~\ref{cfr_spec} (top).
The comparison Is made on the basis of the best match to the $+16$~d and the $+24$~d spectra with those of other CCSNe obtained using the \textsc{gelato\footnote{\url{https://gelato.tng.iac.es/}}} comparison tool \citep{2008A&A...488..383H}.
% Figure 12
A fair agreement is obtained with the Type IIb SN~1993J \citep{1993AAS...183.3903G,1993AAS...183.3902B,1995A&AS..110..513B} and the Type II SNe~2007od \citep{2010ApJ...715..541A,2011MNRAS.417..261I} and 2009bw \citep{2012MNRAS.422.1122I}. 
Nonetheless, we note that a better match is provided by a sample of superluminous SNe \citep[SLSNe;][Figure~\ref{cfr_spec}, bottom]{2012Sci...337..927G}, in particular SN~2008es \citep{2009ApJ...690.1303M,2009ApJ...690.1313G}.
SN~2008es is an over-luminous Type II-L SN at z$=$0.205 with an absolute $R$-band magnitude of $\simeq-22\,\rm{mag}$. 
Although it reached a much higher luminosity, it showed a temperature evolution similar to that observed for LSQ13zm (namely from $\sim14000$~K to $\sim6400$~K during the first 65~d after maximum), along with comparable line velocities ($\simeq10000$\kms) for the broad components. 
The Type II SN CSS121015:004244$+$132827, likely an interacting object \citep{2014MNRAS.441..289B}, has also a similar spectrum. This object has a $B$-band absolute magnitude of $\simeq-22.6\,\rm{mag}$, and shows a linearly declining light-curve.

Among the sample of SNe with H-rich CSM, also the energetic Type IIn SN~2008fz \citep{2010ApJ...718L.127D} provides a good match with our spectrum of LSQ13zm.
SN~2008fz reaches an absolute $V$-band magnitude of $\simeq-22\,\rm{mag}$ with an inferred radiated energy of $\gtrsim1.4\times10^{51}$~\ergs, showing slow-evolving light-curves and spectra with multi-components emission lines.
The comparison in Figure~\ref{cfr_spec} (bottom) suggests that all these luminous, CSM-interacting SNe are likely CCSN events. \\

\begin{figure}
\includegraphics[width=1.0\columnwidth]{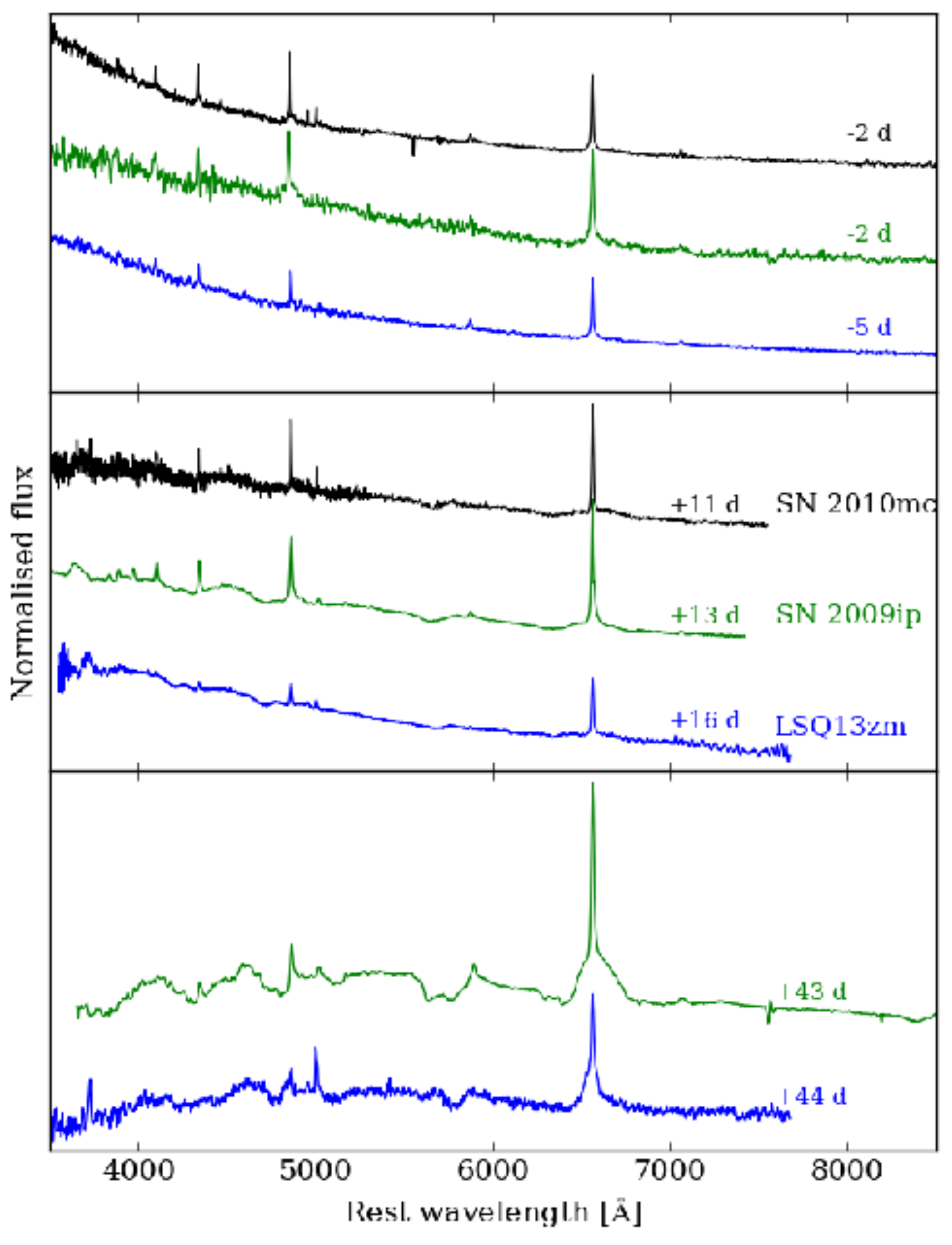}
\caption{Comparison of the spectra of LSQ13zm, SN~2009ip and SN~2010mc at similar phases. The phases of SN~2009ip refer to the 2012b maximum, those of SN~2010mc to the maximum light.}
\label{10mc_compare}
\end{figure}

In Figure~\ref{10mc_compare}, we also report a comparison of the $-$5~d, $+$16~d and $+$44~d spectra of LSQ13zm (2013b event), with the spectra of SN~2009ip \citep{2013ApJ...767....1P} and SN~2010mc \citep{2013Natur.494...65O} at similar phases.
These two comparison objects are characterised by a similar sequence of events as those observed in LSQ13zm. 
The strong similarity in the spectroscopic and photometric properties of the three objects is remarkable, although this does not necessarily imply that their sequence of photometric outbursts has to be interpreted in a similar way (see Section~\ref{discussion}).\\

 % Figure 13

%%%%%%%%
% SECTION 5
%%%%%%%%
\section{Dating mass-loss episodes} \label{preSN}
In Section~\ref{spectroscopy}, we reported the results of our spectroscopic follow-up campaign of LSQ13zm.
A blue continuum with prominent narrow Balmer lines in emission characterises the spectra at around the 2013b maximum, suggesting the presence of a photo-ionised un-shocked CSM, which recombines and re-emits photons.
If the 2013b event was a genuine SN explosion, one might claim that this ionised gas was expelled during the 2013a outburst. \\

The spectra show the first unequivocal signs of interaction (namely the rise of the intermediate \halpha~component) between $+44$~d and $+67$~d after maximum. 
The onset of the interaction, in fact, is generally characterised by the emergence of intermediate-width components (i.e. with FWHM velocities a few $10^3$\kms) and an increase in the continuum luminosity of the transient.
We remark that the absence of a significant increase in the spectral continua between $+44$~d and $+67$~d is probably the consequence of the strong contamination by the host galaxy starting at these phases.
Hereafter, we will assume that the SN explosion occurred 16~d before maximum ($\rm{MJD}_{expl}=56387.9\pm2.5$, computed from a parabolic fit to the $R$-band light-curve of the 2013b event). 
Considering an intermediate phase (namely day 55 post-max) as the indicative epoch when the SN ejecta reach the pre-existing CSM, we can approximately estimate the epoch of the ejection of this circumstellar shell.
Adopting 11500\kms~as an indicative velocity for the SN ejecta (the velocity inferred from the broad component of the $+8$~d spectrum) and $470$\kms~as the velocity of the un-shocked CSM (as inferred from our highest resolution spectrum), we find that this gas was expelled by the star $\simeq4$~yr before the putative SN explosion. 
Therefore this material had been lost well ahead the 2013a event.
Under the same gross assumptions, we can infer that the gas moving at 470\kms~and expelled at the beginning of the 2013a event (that we assume to be on $\rm{MJD}_{out}=56366.1\pm3.0$, i.e. $\sim22$~d before the epoch of the explosion; also estimated through a polynomial fit of the $R$-band light-curve) has been blown away by the SN ejecta in $\sim1$~d. \\

The SN shock breakout and the early interaction between SN ejecta and inner CSM would provide enough energy to ionise the outer CSM expelled in past, unobserved mass-loss events. The latter mechanism would the most natural explanation for injecting ionizing radiation to power the narrow \halpha~line produced in the outer CSM, and observed from our first spectrum of LSQ13zm to that at $\sim44$~d after light curve peak. 
We note that some spectroscopic indicators would argue against strong ejecta-CSM interaction at early phases in LSQ13zm. In fact, until day $+44$, there is no direct evidence for the presence intermediate-width line components typical of shocked gas shells, and broad features from un-shocked SN ejecta are clearly detected.
Nonetheless, higher-density clumps in more diluted gas and/or geometrical effects may produce the above mentioned spectral observables. \\

At epoch later than $+44$~d, the interaction between the ejecta of LSQ13zm and outer CSM becomes more evident, with the detection of intermediate-width line components. 
This provides a sufficient amount of high-energy photons to ionise the un-shocked external material, as proposed by \citet{1994MNRAS.268..173C} to explain the observed spectroscopic features of SN~1988Z.
We will widely discuss the implications of the photometric and spectroscopic properties of LSQ13zm in Section~\ref{discussion}.

%%%%%%%%
% SECTION 6
%%%%%%%%
\section{On the nature of LSQ13zm} \label{discussion}
In Sections~\ref{photometry} and \ref{spectroscopy}, we reported the photometric and spectroscopic analysis of the optical transient LSQ13zm.
In a remarkable sequence of events, this object showed a first outburst, the `2013a event', during which it reached the absolute magnitude $M_R=-14.87\pm0.25\,\rm{mag}$.
This brightening was followed after $\sim3$ weeks by another episode, the `2013b event', reaching an absolute peak magnitude $M_R=-18.46\pm0.21\,\rm{mag}$, similar to those observed in Type IIn SNe.
We need to remark that the presence of previous outbursts in the past decade cannot be ruled out since, as shown in Figure~\ref{abslc}, our historical limits are not deep enough to detect eruptive episodes fainter than $M_R\simeq-15\,\rm{mag}$.
Moreover, as discussed in Section~\ref{preSN}, the results of our photometric and spectroscopic analysis suggest that an unobserved eruptive event might have occurred $\gtrsim4$~years before the 2013a event. \\

Our spectra, covering almost 2 years after the 2013b episode onset, are characterised by three main phases.
At early phases (before and around the 2013b maximum), the spectra show a blue continuum (with an inferred black-body temperature of $\sim13000$~K) with narrow Balmer lines in emission characterised by Lorentzian profiles.
At intermediate epochs (namely during the early decline after the 2013b light-curve maximum), the spectra show broad absorption features, particularly evident for the \halpha, \hbeta~and \ion{He}{I}/\ion{Na}{ID} features, suggesting the presence of underlying high-velocity ejecta.
Finally, late-phase spectra are dominated by the intermediate-width features typical of ejecta-CSM interaction, although the contamination of the host galaxy strongly affects the spectral appearance at these phases. \\

While early-phase spectra show unequivocal evidence of a surrounding photo-ionised CSM, different physical mechanisms can produce the ionising photons required to explain the observed features.
Lorentzian wings are, in fact, typical of hot ionised gas and are usually related to Thomson-scattering due to free electrons in the medium.
While inner ejecta-CSM collisions may produce the sufficient amount of energy to ionise pre-existing H-rich material, narrow recombination lines may also be powered by a long-lived shock-breakout within a dense, optically thick surrounding CSM.
The lower velocity limit for SN shocks breaking out in such dense media was found to be $\sim10^4$\kms, supporting the conclusion that the rise of the light-curves of some Type IIn SNe might be powered by shock break-out within a dense CSM \citep{2014ApJ...788..154O}.
\cite{2010ApJ...724.1396O} also explained the fast rising UV emission with shock break-out through a dense CSM for the Type IIn SN PTF09uj.
The subsequent visible emission at later times was then interpreted as the diffusion of the energy deposited in the CSM by the shock itself. \\

The spectroscopic evolution during the early decline phases is characterised by prominent absorption features strengthening with time, with relatively high and constant expansion velocities ($\sim10^4$\kms, inferred from the minima of the absorption profiles).
The absorption feature observed for \halpha~and \ion{He}{I}/\ion{Na}{ID} is of particular interest, showing boxy profiles with wing velocities extending to $\sim22000$\kms~(see Figure~\ref{hhedetail}), without showing a significant evolution from $\sim+8$~d to $\sim+44$~d after maximum.
Such very high velocities, along with the peculiar profile of \halpha~and \ion{He}{I}/\ion{Na}{ID}, were never observed in SN~2009ip, which, on the other hand, showed blue wings with an inferred velocity of $\sim13000$\kms~only. \\

Clear signatures of ejecta-CSM interaction become unequivocally visible only at late phases, after $+67$~d, when an intermediate component clearly appears in the \halpha~line profile.
Moreover, in analogy with many other interacting transients, the late spectra show no trace of $\alpha$- or Fe-peak elements produced in the stellar/explosive nucleosynthesis. \\

The case of SN~2009ip is widely debated, and different interpretations to the nature of the two 2012 events are offered in the literature.
While the photometric properties of the 2012a event were comparable with those displayed by SN impostors, the spectroscopic analysis at these epochs showed broad P-Cygni line components.
However, broad absorptions had already been observed during the numerous re-brightenings in the 2009-2011 period, and this was an indication that a relatively small amount of gas was expanding at velocities much higher than those typically observed in erupting LBVs. \\

The effects of a particular geometrical configuration on the observables were discussed by \cite{2014MNRAS.442.1166M} using spectropolarimetry data of SN~2009ip.
Their analysis revealed that the two eruptive episodes occurred in 2012 were both highly aspherical (with the second episode showing an higher level of asphericity) and, most importantly, exhibited orthogonal geometries on the sky.
Their results supported the scenario according to SN~2009ip exploded during the 2012a event with a prolate/bipolar photosphere geometry partially obstructed by a toroidal (disc- or ring-shaped) dense CSM \citep[see Figure 9 in][]{2014MNRAS.442.1166M}.
Toroidal geometry for the surrounding CSM of SN~2009ip was already suggested by \cite{2014AJ....147...23L} on the basis of spectroscopic arguments, while bipolar geometry was proposed by \cite{2001ApJ...550.1030W} to explain other cases of highly polarised CCSNe.
We cannot rule out that high asphericity can characterise also LSQ13zm, although we do not have spectropolarimetry observations to verify this claim. \\

In the forthcoming sections, we compare the photometric and spectroscopic properties of LSQ13zm with those of other similar transients, in particular SN~2009ip, in order to infer the true nature of the 2013a,b events.
To do that, we consider for LSQ13zm the most plausible scenarios on the basis of the different interpretations given for SN~2009ip.
In Section~\ref{pulspairinst} we will discuss the scenario proposed by \citet{2013ApJ...767....1P}, which supports a pulsational pair-instability (PPI) event powered by collisions between shells expelled at different times, while in Section~\ref{interaction} we will inspect the scenario proposed by \citet{2013MNRAS.430.1801M}, suggesting a SN explosion followed by ejecta-CSM interaction. In Section~\ref{mergers} we will briefly analyse the possibility that the 2013a event is the result of binary interactions between an evolved LBV star with a un-evolved, less massive star, with the 2013b event triggered by a violent merger \citep[following the interpretation proposed by][for SN~2009ip]{2013ApJ...764L...6S,2013MNRAS.436.2484K}. 
Finally, in Section~\ref{explosion} we will discuss the scenario in which the giant eruption of a massive progenitor (most likely an LBV star) is followed by a terminal explosion. \\

% Figure 14
\begin{figure} 
\includegraphics[width=1.0\columnwidth]{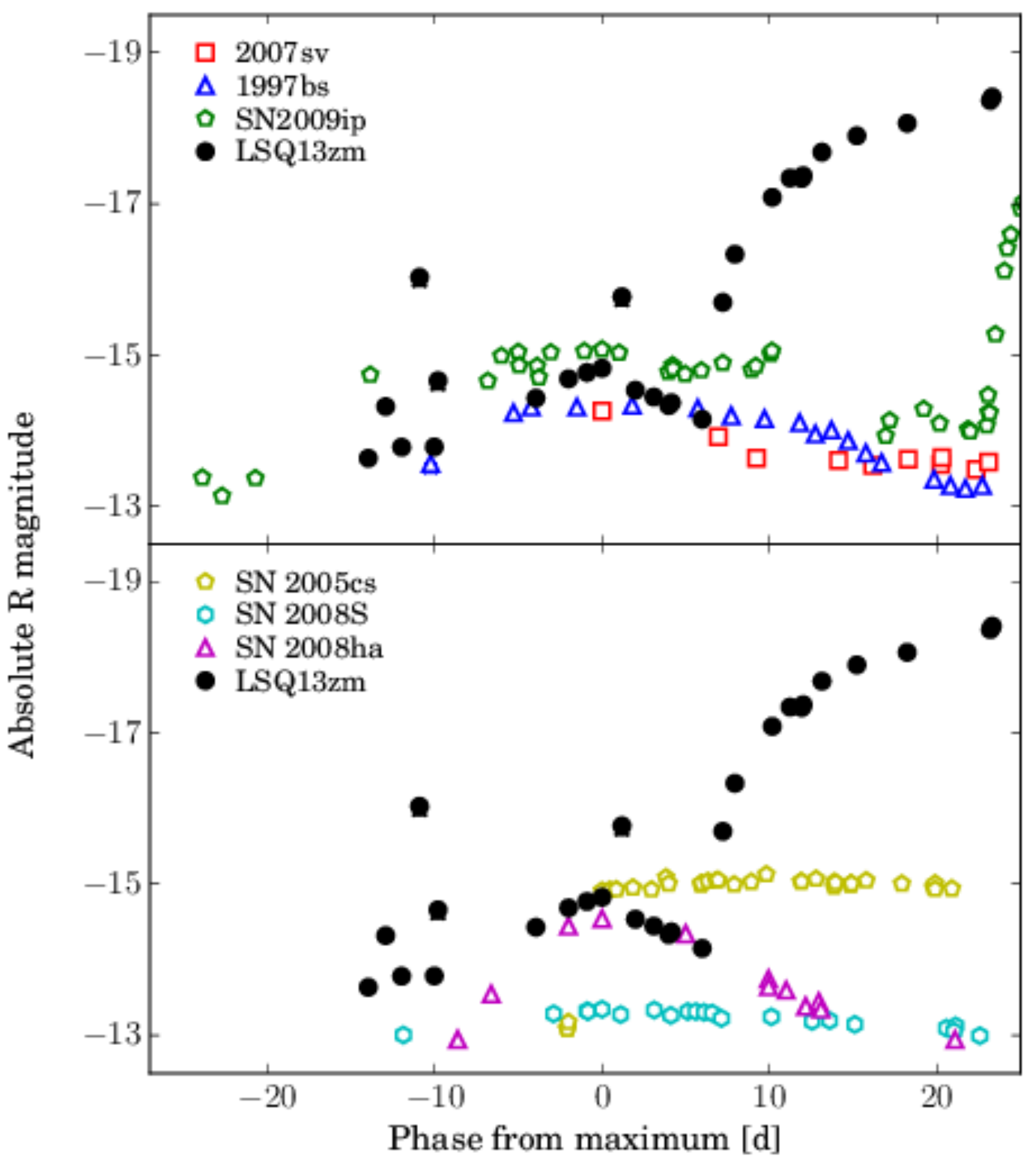}
\caption{{\bf Top:} Comparison of the absolute $R$-band light curve of LSQ13zm with those of known SN impostors. The distance moduli of 2007sv ($\mu=31.38\,\rm{mag}$), 1997bs ($\mu=31.1\,\rm{mag}$) were taken from \citet{2015MNRAS.447..117T} and \citet{2000PASP..112.1532V} respectively, while the reddening estimates for the same two objects ($A_V=0.056\,\rm{mag}$ and $A_V=0.093\,\rm{mag}$ respectively) were taken from the NED archive. The phases are relative to the maximum light. For LSQ13zm we refer to the maximum of the 2013a episode, for SN 2009ip to the maximum of the 2012a event \citep{2013ApJ...767....1P}. {\bf Bottom:} Comparison with the absolute $R$-band magnitudes of a sample of faint SNe. The distance moduli and the reddening estimates for SN~2005cs ($\mu=29.26\,\rm{mag}$, $A_V=0.155\,\rm{mag}$), SN~2008S ($\mu=28.74\,\rm{mag}$, $A_V=1.13\,\rm{mag}$) and SN~2008ha ($\mu=31.55\,\rm{mag}$, $A_V=0.236\,\rm{mag}$) and they were taken from \citet{2009MNRAS.394.2266P,2009MNRAS.398.1041B} and \citet{2009Natur.459..674V} respectively. The phases are relative to the maximum light. Phase 0 in LSQ13zm is coincident with the epoch of the 2013a event maximum. }
 \label{cfr_3pan}
\end{figure}

%%%%%%%%%
% SECTION 6.1
%%%%%%%%%
\subsection{2013b as a pulsation pair-instability event powered by shell-shell collisions} \label{pulspairinst}
The first scenario illustrated for LSQ13zm is that of an unusual SN impostor, where the first, weaker light curve peak would be produced by a giant, non-terminal outburst, followed by a major re-brightening due to 
interaction between two dense circumstellar-shells. To support this, we first compare in Figure~\ref{cfr_3pan} (top panel) the absolute light curve of the LSQ13zm 2013a event with those of proposed SN impostors, and with the 2012a event of SN~2009ip. All these transients show very similar absolute peak magnitudes, which are relatively faint for SNe (but see Figure~\ref{cfr_3pan}, bottom panel, and Section~\ref{interaction}), and frequently observed in non-terminal eruptions of massive stars \citep[see, e.g.][and references therein]{2011MNRAS.415..773S}. What triggers these eruptions is still debated. One possibility is that `pulsation pair-instability' (PPI) in very massive stars may produce some of these impostors. \\

During a PPI event, the production of electron-positron pairs due to high temperatures leads to a reduction of the radiative pressure inside the core followed by a partial collapse and runaway thermonuclear explosions.
However, these explosive events have not necessarily sufficient energy to unbind the entire star causing the disruption of the progenitor, but can provide a sufficient amount of energy to trigger violent ejections of several solar masses of the envelope.
If the remaining core mass is high enough, subsequent eruptions may happen, and collisions of shells expelled at different times would eventually lead to a dramatic increase in the luminosity of the transient.
The conversion of kinetic energy into radiation can be extremely efficient, providing an energy output up to $10^{50}$~erg, much greater than that measured even in CCSNe.
For this reason, PPI is one of the mechanisms proposed to explain the major eruption of LBVs \citep{2007Natur.450..390W}. \\

In a detailed spectroscopic analysis, \citet{2013ApJ...767....1P} compared the spectra of SN~2009ip collected during the erratic variability in 2009-2011, and during the 2012a episode with those of the SN impostor NGC~3432-LBV1 \citep[aka 2000ch;][]{2004PASP..116..326W,2010MNRAS.408..181P}, very likely an hyper-active LBV.
In their paper, Pastorello et al. showed that high velocity ejecta can be observed also during major eruptions of LBVs, with broad wings extending up to 9000\kms. 
As the erratic variability observed in SN~2009ip from 2009 to 2011 closely resemble that still experienced by NGC~3432-LBV1, one of the main conclusions of \citet{2013ApJ...767....1P} was that the major re-brightening of SN~2009ip (called 2012b) was the consequence of a PPI event, in which the gas expelled in a major outburst (possibly the event 2012a) collided with pre-existing CSM, likely collected during the previous erratic variability phase of the massive LBV precursor. 
The consequence was that SN~2009ip did not undergo core-collapse during the 2012b event, but was the result of collisions among shells expelled at different times.
Multiple repeated outbursts, similar to those observed in SN impostors and not leading to a terminal SN explosion are a possible scenario, and have already been observed (see e.g. the case of SNHunt248; \cite{2015A&A...581L...4K,2015MNRAS.447.1922M} or SN~1994W; \cite{2009MNRAS.394...21D}). 
According to the above scenario, the 2013a event in LSQ13zm would be a major outburst produced by PPI, while the light-curve of the 2013b event would be a large re-brightening powered by the conversion of kinetic energy into radiation. 
With the above scenario, there would be no need of additional energy input from the decay of radioactive material. \\ 

Nonetheless, the PPI plus shell-shell collision does not convincingly explain all the observables of LSQ13zm.
First of all, some concerns may derive from energetic consideration.
As we do not have any data in the ultraviolet (UV) domain for LSQ13zm, we estimate its quasi-bolometric peak luminosity assuming a UV contribution to the total luminosity similar to that of SN~2009ip. 
We obtain a peak luminosity for LSQ13zm of $\gtrsim10^{43}$~\ergs.
Even assuming a 10\% of efficiency, the conversion of the kinetic energy of 1~\msun~of gas moving at 11500\kms~(a crude estimate of the velocity of the ejecta based on our early spectra of LSQ13zm) into radiation would give $\sim10^{51}$~erg of total radiated energy, which is a factor $\sim10$ larger than the total energy radiated during the 2013b event.
We also remark that this value might even be significantly underestimated since, as pointed out by \cite{2014MNRAS.442.1166M}, particular geometrical configurations may lead to a reduction in the ejecta-CSM interacting surface and, hence, a less efficient conversion of kinetic energy into radiation.
However, the spectral properties of LSQ13zm, including the presence of very broad features similar to those observed in CCSNe spectra (see Figure~\ref{cfr_spec} and discussion in Section~\ref{spec_compare}), do not favour the PPI mechanism for the 2013a event, favouring an alternative explanation (Section~\ref{explosion}).

%%%%%%%%%
% SECTION 6.2
%%%%%%%%%
\subsection{The 2013a event was a faint SN, with 2013b being powered by ejecta-CSM interaction} \label{interaction}
Another possibility is that we first observed a very weak CCSN followed by the interaction between the ejecta and pre-existing CSM.
This scenario was first proposed by \citet{2013MNRAS.430.1801M} to explain the chain of events producing the final brightenings of SN~2009ip,
and should be re-discussed in the context of LSQ13zm.
The faint absolute magnitude of the 2013a event is not a major problem, since weak CCSNe do exist (e.g., Figure~\ref{cfr_3pan}, bottom panel).
Weak explosions may result from the CC of moderate-mass (8-10~\msun) super--AGB stars producing so-called `electron-capture (EC) SNe' \citep[e.g.][]{2009ApJ...705L.138P}, as well as from the core-collapse of more massive stars (above 25-30~\msun) where a large fraction of the progenitor's mantle falls back onto the nucleus, likely producing a black hole \citep{2003ApJ...591..288H}.
In both terminal explosion scenarios, the final outcomes are expected to be faint CCSNe with absolute magnitudes ranging from $-13\,\rm{mag}$ to $-15\,\rm{mag}$. \\

Although faint absolute magnitudes at maximum are often arguments used to discriminate SN impostors from genuine SNe \citep[see e.g.][]{2015MNRAS.447..117T}, as pointed out by \citet{2007Natur.449E...1P} there are a few classes of genuine SNe characterised by weak peak magnitudes, including the most common CCSNe, i.e. Type II-P events. 
The faintest member of this class was SN~1999br \citep{2004MNRAS.347...74P}, reaching an absolute peak magnitude of $\simeq-14.2\,\rm{mag}$, but also SN~2010id showed a similarly faint magnitude \citep[$M_R<-14$mag;][]{2011ApJ...736..159G}. 
This evidence was also used by \citet{2013MNRAS.430.1801M} and \citet{2014MNRAS.438.1191S} to state that the weakness of the 2012a event in SN~2009ip was not a strong argument to rule out the CCSN scenario. \\

Consequently, the photometric analysis alone is not sufficient to rule out either the genuine SN or the SN impostor scenarios for this class of transients \citep{2015MNRAS.447..117T}. 
Therefore, additional clues are needed to discriminate between the different types of explosions. 
One plausible clue is the width of the broadest line components, that can be associated with the velocities of the fastest-moving ejected material. In the spectra of SN~2009ip, the detection of broad \halpha~line components led \citet{2013MNRAS.430.1801M} to conclude that it was indeed the final explosion of an LBV progenitor, although this argument has been questioned by other authors, \citep[e.g.,][]{2013ApJ...767....1P,2015MNRAS.453.3886F,2014ApJ...780...21M}. \\

Following \citet{2013MNRAS.430.1801M}, in Figure~\ref{cfr_3pan} (bottom panel) we compare the absolute light-curve of the 2013a event in LSQ13zm with those of a sample of faint SNe, including the Type II-P SN~2005cs \citep{2006MNRAS.370.1752P}, the peculiar Type IIn SN~2008S \citep{2009MNRAS.398.1041B} and SN~2008ha \citep{2009Natur.459..674V}. 
Clearly, the absolute magnitude and the evolution of the light-curve of the 2013a event of LSQ13zm are consistent with those of faint SNe. 
For this reason, assuming that 2013a is a true SN explosion, we then need to provide a coherent interpretation for the 2013b event.
As mentioned in the previous section, we consider the possibility that the 2013b re-brightening is powered by the interaction of the SN-ejecta with a dense CSM expelled during past mass loss episodes. \\

Collisions between H-rich SN ejecta and circum-stellar gas usually produce spectra with a blue continuum and strong Balmer lines in emission, with prominent intermediate-width components (namely with FWHM velocities of a few $\sim10^3$\kms).
Intermediate-velocity components are typical of the spectra of interacting objects, and arise in the gas interface between the forward and reverse shocks \citep[see][]{1994ApJ...420..268C}. 
However, the first two spectra in Figure~\ref{specSeq} are characterised by a blue continuum with narrow H and \ion{He}{I} lines in emission, with low FWHM velocities (a few $10^2$\kms), and no evidence of intermediate components.
Narrow lines are generally indicative of a slow-moving, ionised thin shell, that may eventually produce an opaque pseudo-photospere masking the underlying ejecta-CSM interaction.
After the 2013b event peak, the spectra evolve showing very broad features similar to those of canonical non-interacting CCSNe, and likely attributed to SN ejecta.
The fact that the spectral signatures can now be directly observed, can be explained with a peculiar geometrical configuration of the CSM \citep[e.g.][]{2014MNRAS.442.1166M}, or a non-homogeneus clumpy CSM structure.
The characteristic boxy absorptions detected in the \halpha~and \ion{He}{I}/\ion{Na}{ID} regions up to $\sim+44$~d from the 2013b maximum, suggest the presence of a fraction of expelled gas moving at very high velocities (up to 22000\kms).
Adopting a scenario according to which the 2013a event is a faint SN followed by ejecta-CSM interaction (2013b), the very high expansion velocities would be measured at over 90~d after the explosion.
Such high velocities have never been observed at late SN phases in any CCSN Type.
Hence this observed parameter would not favour the SN scenario for the 2013a event of LSQ13zm.
 
%%%%%%%%%
% SECTION 6.3
%%%%%%%%%
\subsection{2013a and 2013b events generated by repeated binary interactions and a final merger-burst} \label{mergers}
A further scenario that could potentially explain the unusual light curve of LSQ13zm was first proposed for SN~2009ip by \cite{2013ApJ...764L...6S} and \cite{2013MNRAS.436.2484K}.
An overall similarity, in fact, should be remarked between the light curves of SN~2009ip (and - consequently - LSQ13zm) and the binary merger V838~Mon (\citealt{2002IAUC.7785....1B,2002MNRAS.336L..43K,2002A&A...389L..51M,2003MNRAS.341..785C,2003ApJ...582L.105S,2004A&A...416.1107K,2005A&A...436.1009T}, but see \citealt{goranV838mon} for a different interpretation on the progenitor system).
According to this scenario, the sequence of outbursts experienced by SN~2009ip before the main 2012b event, is explained with mass transfer in a strongly interacting binary system consisting in an evolved massive ($M_1=60$-100\msun) star with a lower-mass main-sequence companion of $M_2=12$-50\msun~lying in an eccentric orbit. A final merger-burst would be responsible for the re-brightening observed during 2012b.
A binary merger \citep[e.g.,][]{2006MNRAS.373..733S} is a violent event in which two (or more) stars merge, due to direct collision or interaction.
As the primary star evolves beyond the main sequence, it expands triggering the merging event, although also orbital angular momentum losses via tidal interaction can lead the binary system to a collision.
Extreme cases of mass transfer at very high rates have also been suggested as an alternative mechanism to trigger outbursts or eruptive episodes of evolved massive stars like LBVs \citep[see e.g.][]{2010ApJ...723..602K}. 
The presence of multiple minor peaks in the light-curve of SN~2009ip after the 2012b maximum, carefully analysed by \citet{2015AJ....149....9M}, is in fact reminiscent of the secondary luminosity peaks observed during the Giant Eruption of $\eta$~Car, and attributed to binary interaction at periastron passages \citep{1996ApJ...460L..49D,2010ApJ...723..602K,2011MNRAS.415.2009S}.
These bumps, superposed to the main light curve of the 2012b event, were observed with amplitudes of several tenths of magnitude \citep{2015AJ....149....9M}. 
\citet{2013ApJ...764L...6S} and \citet{2013MNRAS.436.2484K} focused their discussion, in particular, on one major fluctuation, which occurred at $+30$~d after the 2013b maximum.
In their scenario, this peak is due to interaction of the gas expelled during the merger event, with a low mass shell expelled in a previous binary interaction episode. \\

As LSQ13zm, SN~2009ip and (more marginally) V838~Mon show similar light-curves, the above scenario has to be considered also to explain the photometric properties of LSQ13zm.
While erratic variability has been registered for SN~2009ip in the period 2009-2011 \citep{2013ApJ...764L...6S}, these were not detected in the historical data of LSQ13zm.
However, as mentioned in Section~\ref{photoanalysis}, these detection limits were not deep enough to rule out that outbursts fainter than $-15$ mag had occurred in the past, including eruptive mass-loss events triggered by binary interaction.
On the other hand, luminosity fluctuations similar to those observed during the 2012b event in SN~2009ip are not observed during the 2013b event of LSQ13zm, whose light curve shows a monotonic decline in all bands. 
For this reason, the binary merger scenario is less plausible for LSQ13zm, although we admit that our data are not sufficiently well-sampled to definitely rule out this scenario.

%%%%%%%%%
% SECTION 6.4
%%%%%%%%%
\subsection{2013a was an eruption, 2013b the SN explosion} \label{explosion}
The most promising scenario which explains fairly well the double brightening observed in the light curve of LSQ13zm includes an initial outburst similar to those observed in some LBVs (i.e. a SN impostor), followed by the genuine SN explosion a short time later.
The photometric similarity of the 2013a event with light-curves of SN impostors, viz. 1997bs \citep{2000PASP..112.1532V} and 2007sv \citep{2015MNRAS.447..117T} has been remarked in Section~\ref{pulspairinst} (see also top panel in Figure~\ref{cfr_3pan}), and provides support to the non-terminal burst for the first brightening of LSQ13zm. 
However, we have to remark that the nature of the 1997bs transient was recently questioned by \citet{2015MNRAS.452.2195A}, who proposed it to be a faint SN on the basis of new analysis performed on both space and ground-based data. \\

SN impostors are believed to be stellar outbursts with light-curves similar to those of genuine Type IIn SNe, though (on average) with fainter peak luminosities.
As discussed in Section~\ref{intro}, their spectra share some similarity with those of Type IIn SNe, although they usually display a faster evolution in the continuum temperature, and weak or no evidence of multicomponent line profiles (in particular, very broad line wings).
In addition, strong high-velocity $\alpha$-- or Fe-peak element lines typically identified in the spectra of CCSNe are not comfortably detected in those of SN impostors. 
Therefore, the spectra and the photometric data together give crucial information to discriminate between SNe and SN impostors.\\

Since we do not have any direct information on the spectroscopic properties of LSQ13zm during the 2013a event, we have to extrapolate useful information from the available light curve and the spectroscopic features of the 2013b event.
If we assume that 2013b was the real SN light curve, 2013a was necessarily a pre-SN outburst. 
As mentioned above, a conclusive proof for establishing the SN nature of the 2013b brightening would be the detection of $\alpha$-- or Fe-peak nebular lines in the late spectra. However, the late spectra are contaminated by strong residual features from foreground emission and heavily affected by features produced in the ejecta-CSM interaction.
This is a common issue for many interacting transients, since the signatures of the interaction usually dominate the spectral emission for months to years after the explosion, veiling the typical features of the SN ejecta during the nebular phase.
Interestingly, from $+16$~d past-maximum, a broad \halpha~emission component and other shallower spectral features with blue-shifted P-Cygni components (cfr. Section~\ref{linesIDspecEV}) become prominent, and are interpreted
as the first direct evidence of the underlying SN ejecta.
The broad component of \halpha~dominates the total line flux until $+67\,\rm{d}$, when the intermediate component become preponderant over the others. 
The high photospheric velocities and the ions producing the broad lines in the spectrum of LSQ13zm, are normally observed in the spectra of H-rich CCSNe 
(see the excellent match of the $+16\,\rm{d}$ spectrum of LSQ13zm with that of SN~2008es in Figure~\ref{cfr_spec}). \\

As a natural consequence, if the 2013b event of LSQ13zm is the real SN explosion, as suggested by the detection of broad P-Cygni lines in the spectra taken during that episode, then 2013a should be regarded as the last burp of the progenitor. In other words, the 2013a event is a luminous outburst similar to those observed in active LBVs. 
This conclusion is also supported by the similarity of the light-curve of the 2013a event with those of some SN impostors (Figure~\ref{cfr_3pan}, top.)

%%%%%%%%
% SECTION 7
%%%%%%%%
\section{Conclusions} \label{conclusions}
We presented our study on the photometric and spectroscopic evolution of the optical transient LSQ13zm, that exploded in the galaxy SDSS~J102654.56$+$195254.8, most likely a BDCG with low mass (8.65 to 8.8~Log(M/\msun)) and a relatively low global SFR (0.025~\msun~$\rm{yr}^{-1}$).
The transient object experienced two main brightenings: the former -- labelled as 2013a -- reached an absolute magnitude of $M_R=-14.87\pm0.25\,\rm{mag}$, and the latter -- named 2013b -- with a peak magnitude of $M_R=-18.46\pm0.21\,\rm{mag}$.
The multi-band light curves indicate that LSQ13zm was photometrically similar to the debated SNe~2009ip and 2010mc. 
In particular, the photometric properties (peak magnitude and colours) of the 2013b event of LSQ13zm are consistent with those of a true SN.
On the other hand, the light-curve of the 2013a event is strikingly similar to outbursts of massive stars known as SN impostors, including its faint absolute magnitude at peak and the rapid luminosity decline. \\

The spectroscopic properties of the transient observed during the 2013b episode (namely, the presence of high velocity ejecta with wings up to $\simeq20000$\kms), support the terminal SN explosion, most likely occurred during the 2013b event.
Hence, the 2013a event is an outburst of a massive star, likely an LBV, which occurred few weeks before the stellar core-collapse.
Other considerations on the spectroscopic evolution of LSQ13zm suggest that 2013a is likely the last episode of a sequence of mass-loss events that produced as final outcome the observables of a Type IIn SN. \\

The lack of a clear evidence of synthetised $\alpha$- or Fe-peak elements in the spectra of LSQ13zm is a common feature of Type IIn SNe and does not constitute a problem for the SN interpretation for LSQ13zm.
In fact, the presence of an opaque ionised CSM expelled during the latest stages of the progenitor's life frequently masks the spectral signatures of the SN ejecta. \\

Although most observational clues suggest that the 2013b event observed during the LSQ13zm evolution was actually a genuine CCSN explosion, we cannot definitely rule out the alternative scenarios discussed in previous Sections.
We finally remark that - although LSQ13zm and SN~2009ip are photometrically very similar - they display subtle spectroscopic differences, suggesting a somewhat different interpretation for their sequence of outbursts.

\section*{Acknowledgements}
Based on observations made with: \\
The Cima Ekar 1.82~m Telescopio Copernico of the INAF (Istituto Nazionale di Astrofisica) -- Astronomical Observatory of Padova, Italy. \\
The Italian Telescopio Nazionale Galileo (TNG) operated on the island of La Palma by the Fundacion Galileo Galilei of the INAF at the Spanish Observatorio del Roque de los Muchachos of the Instituto de Astrofisica de Canarias. \\
The La Silla Quest (LSQ) ESO 1.5~m Schmidt telescope (ESO La Silla, Chile). \\
The Intermediate Palomar Transient Factory (iPTF) 1.2~m Samuel Oschin telescope. \\
The LCOGT 1.0-m telescope at McDonald Observatory (Texas, USA) and the 2.0-m FTN with FLOYDS of the LCOGT network. \\
The Gran Telescopio Canarias (GTC) operated on the island of La Palma at the Spanish Observatorio del Roque de los Muchachos of the Instituto de Astrofisica de Canarias. \\
The Liverpool Telescope operated on the island of La Palma by Liverpool John Moores University at the Spanish Observatorio del Roque de los Muchachos of the Instituto de Astrofisica de Canarias with financial support from the UK Science and Technology Facilities Council. \\
The Nordic Optical Telescope (NOT), operated by the NOT Scientific Association at the Spanish Observatorio del Roque de los Muchachos of the Instituto de Astrofisica de Canarias. \\
The William Herschel Telescope (WHT) operated on island of La Palma by the Isaac Newton Group at the Spanish Observatorio del Roque de los Muchachos of the Instituto de Astrofisica de Canarias. \\
The 6~m Bolshoi Teleskop Alt-azimutalnyi (BTA) and the 1~m Zeiss-1000 telescope (both located at Mount Pastukhov, Caucasus Mountains, Russia). \\
The 0.6~m Rapid Eye Mount (REM) telescope (ESO La SIlla, Chile). \\
The Catalina Real Time Survey (CRTS) Catalina Sky Survey (CSS) 0.7~m Schmidt and Mt.~Lemmon Survey (MLS) 1.5~m Cassegrain telescopes. \\
The 2.2~m University of Hawaii telescope. \\
The UIS Barber Research Observatory 20" telescope. \\
The 2.2~m University of Hawaii telescope. \\
We thank C.~Barbieri, G.~Naletto, L.~Zampieri for the support with the observations of LSQ13zm with ACAM and M.~Graham for the LCOGT observations. \\
AP, SB, NER, AH, LT, GT, and MT are partially supported by the PRIN-INAF 2014 with the project `Transient Universe: unveiling new types of stellar explosions with PESSTO'. \\
NER acknowledges the support from the European Union Seventh Framework Programme (FP7/2007-2013) under grant agreement n. 267251 `Astronomy Fellowships in Italy' (AstroFIt). \\
MS acknowledges support from the Royal Society and EU/FP7-ERC Grant No. [615929]. \\
ST acknowledges support by TRR33, "The dark Universe" of the German Research Fundation. \\
AGY is supported by the EU/FP7 via ERC grant No. 307260, the Quantum Universe I-Core program by the Israeli Committee for planning and budgeting and the ISF, by Minerva and ISF grants, by the Weizmann-UK "making connections" program and by Kimmel and ARCHES awards. \\
AMG acknowledges financial support by the spanish Ministerio de Econom\'ia y Competitividad (MINECO), Grant ESP2013-41268-R. \\
VPG, EAB, AFV are supported by Russian Foundation for Basic Research Grant 14-02-00759. \\
SF acknowledges support of the Russian Government Program of Competitive Growth of Kazan University. \\
SS acknowledges support from CONICYT - Chile FONDECYT 3140534, Basal - CATA PFB - 06/2007 and Project IC120009 "Millennium Institute of Astrophysics (MAS)" of Iniciativa Cient\'ifica Milenio del Ministerio de Econom\'ia, Fomento y Turismo.
JM acknowledges the National Science Foundation (NSF, Grant 1108890). \\
REM data were obtained as part of the programme CN2013A-FT-12. \\
ACAM data were obtained as part of the programme OPT 43. \\
The CRTS survey is supported by the US National Science Foundation under grants AST-1313422 and AST-1413600. \\
This research used resources of the National Energy Research Scientific Computing Center, a DOE Office of Science User Facility supported by the Office of the U.S. Department of Energy under Contract No. DE-AC02-05CH11231. \\
A portion of this work was carried out at the Jet Propulsion Laboratory under a Research and Technology Development Grant, under contract with the National Aeronautics and Space Administration. Copyright 2015 California Institute of Technology. All Rights Reserved. US Government Support Acknowledged. \\
LANL participation in iPTF is supported by the US Department of Energy as part of the Laboratory Directed Research and Development program. \\
Funding for SDSS-III has been provided by the Alfred P. Sloan Foundation, the Participating Institutions, the National Science Foundation, and the U.S. Department of Energy Office of Science. \\
The 2MASS project is a collaboration between The University of Massachusetts and the Infrared Processing and Analysis Center (JPL/Caltech). Funding is provided primarily by NASA and the NSF. The University of Massachusetts constructed and maintained the observatory facilities, and operated the survey. All data processing and data product generation is being carried out by IPAC. Survey operations began in Spring 1997 and concluded in Spring 2001. \\
This research was supported by the Russian Scientific Foundation (grant No. 14-50-00043).
\textsc{iraf} is distributed by the National Optical Astronomy Observatory, which is operated by the Associated Universities for Research in Astronomy, Inc., under cooperative agreement with the National Science Foundation. \\
This research has made use of the NASA/IPAC Extragalactic Database (NED) which is operated by the Jet Propulsion Laboratory, California Institute of Technology, under contract with the National Aeronautics and Space Administration.

\appendix
\section{[O~III] lines in the spectra of LSQ13zm}
\begin{figure*} 
\includegraphics[width=0.75\linewidth]{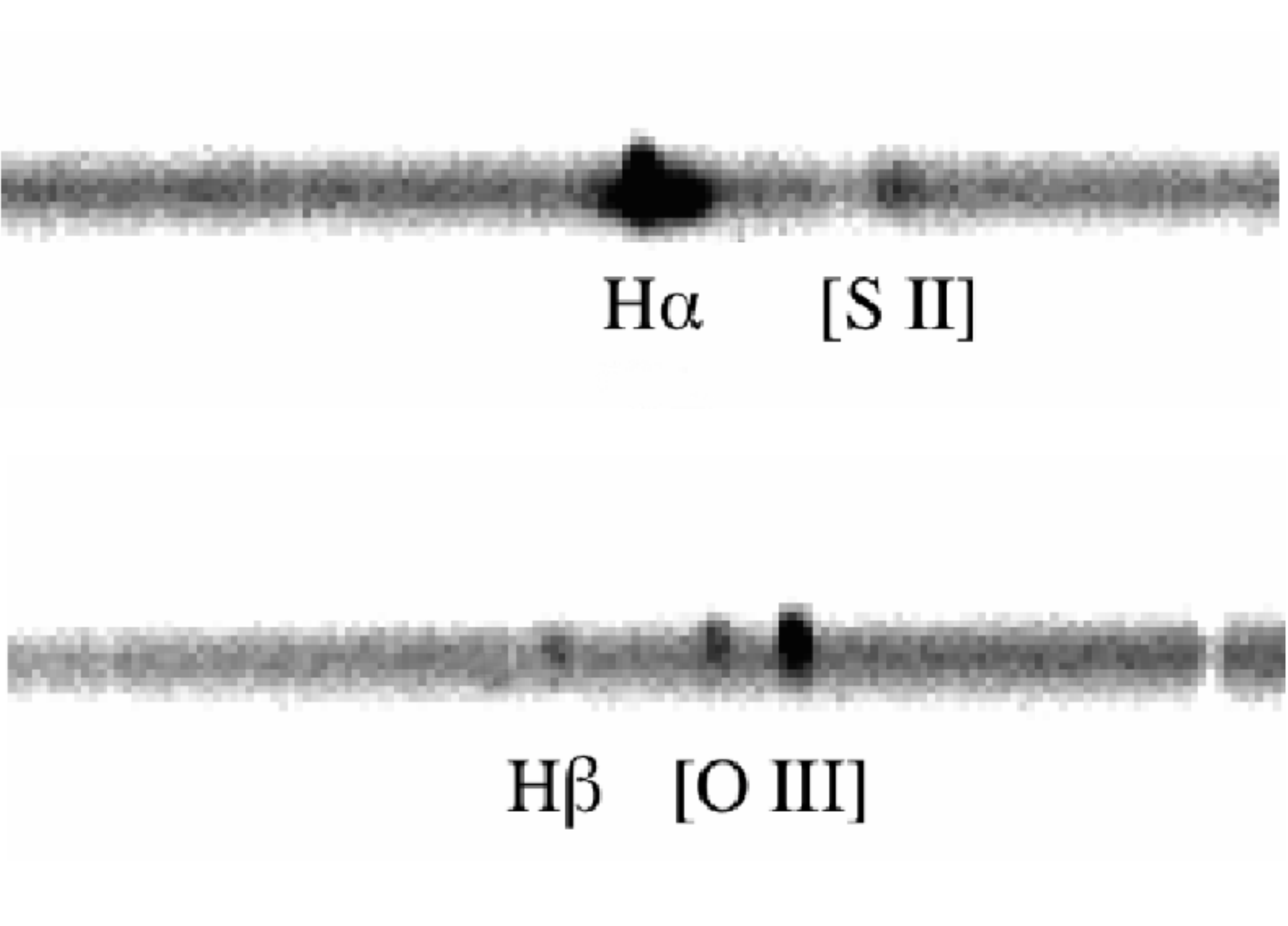}
\caption{Zoom-in of the \halpha~and \ion{O}{III}/\hbeta~regions of the two-dimensional $+$579~d spectrum.}
 \label{spec_2d}
\end{figure*}

Figure~\ref{spec_2d} shows a zoom-in of the \halpha/\ion{S}{II} region and of the \hbeta/[\ion{O}{III}] region in the last $+579$~d two-dimensional spectrum. 
The narrow \halpha~component, as well as the [\ion{O}{III}] doublet, have clearly a broader spatial extension than the SN emission region.
We believe that these lines are not related to the SN CSM, but are part of the strong flux-contamination of the host galaxy.
% Figure 15

\section{Light-curves of LSQ13zm}
\begin{table*}
\begin{minipage}{175mm}                                  
\caption{$griz$ light-curves. The phases are relative to the 2013b maximum.}
\label{grizCurves}  
\begin{tabular}{@{}cccccccl@{}}
\hline
Date & MJD & phase [d] & $g$(err) & $r$(err) & $i$(err) & $z$(err) & Instrument \\
\hline
20130419 & 56401.30 & -5  & 17.40(0.01) & 17.52(0.01) & 17.67(0.01) &     --      & SNIFS     \\
20130424 & 56406.38 & 0   & 17.11(0.05) & 17.22(0.05) & 17.35(0.05) &     --      & SNIFS     \\
20130427 & 56409.24 & 3   & 17.10(0.03) & 17.18(0.03) & 17.32(0.06) & 17.36(0.15) & Spectral Camera    \\
20130428 & 56410.25 & 4   & 17.20(0.02) & 17.18(0.03) & 17.36(0.06) & 17.50(0.09) & Spectral Camera    \\
20130429 & 56411.24 & 5   & 17.17(0.04) & 17.20(0.04) & 17.36(0.06) & 17.54(0.12) & Spectral Camera    \\
20130430 & 56412.24 & 6   & 17.30(0.04) & 17.30(0.05) & 17.42(0.05) & 17.54(0.14) & Spectral Camera    \\
20130502 & 56414.34 & 8   &     --      & 17.34(0.03) & 17.42(0.03) & 17.49(0.30) & Spectral Camera    \\
20130503 & 56415.32 & 9   & 17.57(0.03) & 17.50(0.04) & 17.53(0.03) & 17.70(0.19) & Spectral Camera    \\
20130504 & 56416.27 & 10  & 17.68(0.04) & 17.56(0.04) & 17.62(0.04) & 17.66(0.10) & Spectral Camera    \\
20130520 & 56432.15 & 26  & 19.47(0.06) & 19.09(0.08) & 19.17(0.12) &     --      & SBIG    \\
20130524 & 56436.26 & 30  & 19.50(0.16) & 19.27(0.08) & 19.21(0.07) & 19.35(0.24) & Spectral Camera    \\
20130527 & 56439.35 & 33  & $>19.4$     &     --      & $>19.2$     &     --      & Spectral Camera    \\
20130529 & 56441.15 & 35  & 19.87(0.17) & 19.38(0.08) &     --      &     --      & SBIG    \\
20130529 & 56441.15 & 35  &     --      &     --      & $>19.1$     &     --      & SBIG    \\
20130605 & 56448.16 & 42  & $>19.5$     &     --      &     --      &     --      & SBIG    \\
20130607 & 56450.88 & 45  &     --      &     --      & $>19.4$     &     --      & AFOSC   \\
20130607 & 56450.31 & 44  &     --      & 19.54(0.11) & 19.62(0.16) &     --      & Spectral Camera    \\
20130607 & 56450.31 & 44  & $>19.4$     &     --      &     --      & $>19.0$     & Spectral Camera    \\
20130609 & 56452.30 & 46  &     --      &     --      & $>18.7$     &     --      & Spectral Camera    \\
20130613 & 56456.27 & 50  &     --      & 19.88(0.17) & 19.74(0.15) &     --      & Spectral Camera    \\
20130613 & 56456.27 & 50  & $>19.9$     &     --      &     --      &     --      & Spectral Camera    \\
20130625 & 56468.88 & 63  &     --      &     --      &     --      & 19.78(0.33) & RATCam  \\
20130625 & 56468.88 & 63  &     --      & $>$19.9     & $>19.2$     &     --      & RATCam  \\
20131123 & 56619.17 & 213 &     --      & 22.20(0.35) &     --      &     --      & IO:O    \\
20131125 & 56621.22 & 215 &     --      & 22.23(0.40) &     --      &     --      & IO:O    \\
20131129 & 56625.21 & 219 & 22.54(0.29) & 22.22(0.18) & 21.18(0.16) & 21.57(0.19) & ACAM    \\
20131208 & 56634.20 & 228 &     --      &     --      & 21.38(0.61) &     --      & AFOSC   \\
20131212 & 56638.09 & 232 &     --      &     --      & $>20.2$     &     --      & AFOSC   \\
20140202 & 56690.26 & 284 & 23.16(0.27) & 23.30(0.18) &     --      &     --      & OSIRIS  \\
12341234 & 56688.00 & 282 &     --      &     --      &     --      & 23.65(0.40) & OSIRIS  \\
20140205 & 56693.21 & 287 &     --      &     --      & 23.59(0.12) & 23.79(0.36) & OSIRIS  \\
\hline
\end{tabular}                                                                                                                                            

\medskip
AFOSC: 1.82~m Telescopio Copernico with AFOSC. \\
SBIG: LCOGT 1.0-m telescope at McDonald Observatory (Texas, USA) equipped with an SBIG camera. \\
Spectral Camera: 2.0-m FSTN + FLOYDS. \\ 
SNIFS: 2.2~m telescope of the University of Hawaii with SNIFS. \\
IO:O, RATCam: 2~m Liverpool Telescope with IO:O and RATCam. \\
OSIRIS: 10.4~m Gran Telescopio Canarias (GTC) with OSIRIS. \\
ACAM: 4.2~m William Herschel Telescope (WHT) with ACAM.
\end{minipage}
\end{table*} 
\begin{table*}
\begin{minipage}{175mm}
\caption{$JHK$ light-curves. The phases are relative to the 2013b maximum.}
\label{JHKcurves}
\begin{tabular}{@{}ccccccl@{}}
\hline
Date & MJD & phase [d] & $J$(err) & $H$(err) & $K$(err) & Instrument \\
\hline
20130421 & 56403.99 & -2  & 17.18(0.42) & 16.94(0.28) & $>15.0$      & REMIR  \\
20130423 & 56405.98 & 0   & $>16.1$     & $>$16.2     & $>15.0$      & REMIR  \\
20130425 & 56407.93 & 2   & 16.63(0.07) & 16.34(0.07) & 16.054(0.32) & NOTCam \\
20130426 & 56408.06 & 2   & 16.54(0.30) & 16.28(0.30) & $>15.0$      & REMIR  \\
20130430 & 56412.11 & 6   & 16.77(0.20) & 16.39(0.45) & $>15.0$      & REMIR  \\
20130503 & 56415.07 & 9   &     --      & $>16.4$     & $>14.6$      & REMIR  \\
20130504 & 56416.97 & 11  & 17.38(0.38) & 16.87(0.29) & $>15.6$      & REMIR  \\
20130506 & 56418.04 & 12  & 17.37(0.47) & $>16.9$     & $>15.0$      & REMIR  \\
20130526 & 56438.98 & 33  & 18.10(0.20) & 17.66(0.17) & 17.514(0.15) & NOTCam \\
20130623 & 56466.90 & 61  & 18.91(0.18) & 18.51(0.19) & 18.233(0.13) & NOTCam \\
20131224 & 56650.07 & 244 & 19.06(0.20) & 18.61(0.27) & 18.696(0.25) & NICS   \\
\hline
\end{tabular}

\medskip
REMIR: 0.6~m Rapid Eye Mount (REM) telescope with REMIR. \\
NOTCam: 2.56~m Nordic Optical Telescope (NOT) with NOTCam. \\
NICS: 3.58~m Telescopio Nazionale Galileo (TNG) with NICS.
\end{minipage}
\end{table*}
\clearpage
\newpage
\begin{table*}
\begin{minipage}{175mm}
\caption{$BVR$ light-curves starting from the first 2013a event detection. The phases are relative to the 2013b maximum.}
\label{BVRCurves}
\begin{tabular}{@{}ccccccl@{}}
\hline                                                                                    
Date     & MJD      & phase [d] & $B$(err)    & $V$(err)    & $R$(err)    & Instrument    \\
\hline
20130318 & 56369.11 & -37 &      --     &      --     & 21.79(0.25) & QUEST         \\
20130318 & 56369.16 & -37 &      --     &      --     & $>20.9$     & CFH12K          \\
20130319 & 56370.15 & -36 &      --     &      --     & 21.10(0.65) & SI 600-277    \\
20130319 & 56370.30 & -36 &      --     &      --     & $>21.0$     & CFH12K          \\
20130320 & 56371.10 & -35 &      --     &      --     & 21.64(0.26) & QUEST         \\
20130321 & 56372.17 & -34 &      --     &      --     & $>19.4$     & CFH12K          \\
20130322 & 56373.06 & -33 &      --     &      --     & 21.64(1.01) & QUEST         \\
20130322 & 56373.28 & -33 &      --     &      --     & $>20.8$     & CFH12K          \\
20130328 & 56379.12 & -27 &      --     &      --     & 20.99(0.64) & QUEST         \\
20130330 & 56381.07 & -25 &      --     &      --     & 20.74(0.14) & QUEST         \\
20130331 & 56382.17 & -24 &      --     &      --     & 20.65(0.19) & CFH12K          \\
20130401 & 56383.07 & -23 &      --     &      --     & 20.60(0.14) & QUEST         \\
20130402 & 56384.26 & -22 &      --     &      --     & 19.65(0.44) & SI 600-277    \\
20130403 & 56385.07 & -21 &      --     &      --     & 20.89(0.12) & QUEST         \\
20130404 & 56386.17 & -20 &      --     &      --     & 20.98(0.25) & CFH12K          \\
20130405 & 56387.06 & -19 &      --     &      --     & 21.09(0.15) & QUEST         \\
20130405 & 56387.22 & -19 &      --     &      --     & 21.05(0.26) & CFH12K          \\
20130407 & 56389.05 & -17 &      --     &      --     & 21.27(0.16) & QUEST         \\
20130408 & 56390.29 & -16 &      --     &      --     & 19.73(0.10) & CFH12K          \\
20130409 & 56391.01 & -15 &      --     &      --     & 19.09(0.03) & QUEST         \\
20130411 & 56393.24 & -13 &      --     &      --     & 18.34(0.03) & CFH12K          \\
20130412 & 56394.30 & -12 &      --     &      --     & 18.08(0.02) & CFH12K          \\
20130412 & 56394.99 & -11 &      --     &      --     & 18.09(0.01) & QUEST         \\
20130413 & 56395.08 & -11 &      --     &      --     & 18.05(0.01) & QUEST         \\
20130414 & 56396.21 & -10 &      --     &      --     & 17.74(0.02) & CFH12K          \\
20130416 & 56398.29 & -8  &      --     &      --     & 17.52(0.20) & SI 600-277    \\
20130419 & 56401.30 & -5  & 17.48(0.01) & 17.47(0.01) & 17.36(0.01) & SNIFS     \\
20130424 & 56406.26 & 0   &      --     &      --     & 17.06(0.04) & SI 600-386    \\
20130424 & 56406.38 & 0   & 17.20(0.05) & 17.18(0.05) & 17.01(0.05) & SNIFS     \\
20130501 & 56413.77 & 7   & 17.56(0.03) & 17.38(0.02) & 17.21(0.04) & Zeiss-1000    \\
20130502 & 56414.77 & 8   &      --     & 17.46(0.14) &      --     & SCORPIO       \\
20130502 & 56414.78 & 8   & 17.67(0.03) & 17.45(0.01) & 17.27(0.01) & Zeiss-1000    \\
20130503 & 56415.17 & 9   &      --     &      --     & 17.38(0.21) & SI 600-277    \\
20130504 & 56416.20 & 10  &      --     &      --     & 17.51(0.02) & CFH12K          \\
20130505 & 56417.21 & 11  &      --     &      --     & 17.61(0.02) & CFH12K          \\
20130508 & 56420.13 & 14  & 18.32(0.05) & 18.01(0.04) & 17.83(0.14) & Alta          \\
20130510 & 56422.76 & 16  & 18.42(0.03) & 18.08(0.06) & 18.00(0.25) & SCORPIO       \\
20130521 & 56433.20 & 27  &      --     &      --     & 18.85(0.07) & CFH12K          \\
20130530 & 56442.11 & 36  & $>18.9$     &      --     &      --     & Alta          \\
20130603 & 56446.18 & 40  &      --     &      --     & 19.36(0.21) & SI 600-277    \\
20130607 & 56450.88 & 45  &      --     &      --     & $>19.3$     & AFOSC         \\
20130608 & 56451.79 & 45  &      --     & 20.05(0.05) & 19.51(0.21) & SCORPIO       \\
20130609 & 56452.30 & 46  & 19.74(0.28) &      --     &      --     & Spectral Camera          \\
20130610 & 56453.19 & 47  &      --     &      --     & 19.39(0.00) & SI 600-277    \\
20130627 & 56470.87 & 64  & $>18.2$     &      --     &      --     & RATCam        \\
20131208 & 56634.17 & 228 &      --     &      --     & $>19.9$     & AFOSC         \\
20131212 & 56638.06 & 232 &      --     &      --     & $>20.2$     & AFOSC         \\
20131218 & 56644.17 & 238 & $>20.4$     & $>19.9$     &      --     & IO:O          \\
\hline                                                        
\end{tabular}
                                                     
\medskip
AFOSC: 1.82~m Telescopio Copernico with AFOSC. \\ 
SNIFS: 2.2~m telescope of the University of Hawaii with SNIFS. \\
SCORPIO: 6~m Bolshoi Teleskop Alt-azimutalnyi (BTA) with SCORPIO \\ 
Zeiss-1000: 1~m Zeiss-1000 telescope. \\
QUEST: 60" ESO Schmidt Telescope of the La Silla Quest (LSQ) survey with QUEST. \\
SI 300-386: 1.5~m Cassegrain reflector of the Mt.~Lemmon Survey (MLS) with SI 300-386. \\
SI 600-277: 0.7~m Schmidt telescope of the Catalina Sky Survey (CSS) with SI 600-277. \\ 
CFH12K: 1.2~m Samuel Oschin Telescope of the Intermediate Palomar Transient Factory (iPTF) with CFH12K. \\ 
Spectral Camera: LCOGT 2.0-m Faulkes North Telescope. \\
IO:O: 2~m Liverpool Telescope with IO:O. \\ 
Alta: 20" telescope of the UIS Barber Research Observatory with an Alta U42 CCD Camera.
\end{minipage}
\end{table*}
\clearpage
\newpage
\begin{table*}
\begin{minipage}{175mm}
\caption{$R$-band detection limits of LSQ13zm prior to the 2013a event. The transient was not detected at any reported epoch in the table. The phases are relative to the 2013b maximum.}
\label{photolimits}
\begin{tabular}{@{}ccccc@{}}
\hline
Date  &    MJD   & phase [d] & $R$         & Instrument \\
\hline
20040128 & 53032.26 & -3374 & $>21.2$ & SI 600-277 \\
20050109 & 53379.35 & -3027 & $>20.4$ & SI 600-277 \\
20060104 & 53739.41 & -2667 & $>20.5$ & SI 600-277 \\
20070110 & 54110.46 & -2296 & $>20.9$ & SI 600-277 \\
20070314 & 54173.30 & -2233 & $>20.3$ & SI 600-386 \\
20080111 & 54476.37 & -1930 & $>20.8$ & SI 600-277 \\
20080302 & 54527.40 & -1879 & $>20.1$ & SI 600-386 \\
20081029 & 54768.49 & -1638 & $>20.3$ & SI 600-386 \\
20090102 & 54833.39 & -1573 & $>21.1$ & SI 600-277 \\
20090131 & 54862.31 & -1544 & $>20.5$ & SI 600-386 \\
20090302 & 54892.27 & -1514 & $>20.3$ & SI 600-386 \\
20090517 & 54968.19 & -1438 & $>20.4$ & SI 600-386 \\
20090930 & 55104.53 & -1302 & $>18.7$ & CFH12K     \\
20091003 & 55107.52 & -1299 & $>20.1$ & CFH12K     \\
20091016 & 55120.52 & -1286 & $>20.9$ & CFH12K     \\
20091022 & 55126.49 & -1280 & $>20.9$ & CFH12K     \\
20091031 & 55135.51 & -1271 & $>20.8$ & CFH12K     \\
20091103 & 55138.46 & -1268 & $>20.2$ & CFH12K     \\
20091107 & 55142.45 & -1264 & $>20.6$ & CFH12K     \\
20091116 & 55151.44 & -1255 & $>20.9$ & CFH12K     \\
20091218 & 55183.35 & -1223 & $>20.7$ & CFH12K     \\
20091230 & 55195.50 & -1211 & $>19.7$ & CFH12K     \\
20100106 & 55202.47 & -1204 & $>19.9$ & CFH12K     \\
20100108 & 55204.53 & -1202 & $>18.9$ & CFH12K     \\
20100111 & 55207.48 & -1199 & $>21.4$ & CFH12K     \\
20100115 & 55211.55 & -1195 & $>20.6$ & CFH12K     \\
20100125 & 55221.23 & -1185 & $>20.8$ & CFH12K     \\
20100129 & 55225.29 & -1181 & $>18.3$ & CFH12K     \\
20100209 & 55236.26 & -1170 & $>20.6$ & SI 600-277 \\
20100216 & 55243.54 & -1163 & $>20.9$ & CFH12K     \\
20100217 & 55244.31 & -1162 & $>21.5$ & CFH12K     \\
20100219 & 55246.49 & -1160 & $>20.8$ & CFH12K     \\
20100223 & 55250.53 & -1156 & $>20.7$ & CFH12K     \\
20100224 & 55251.15 & -1155 & $>18.8$ & CFH12K     \\
20100313 & 55268.49 & -1138 & $>20.6$ & CFH12K     \\
20100315 & 55270.18 & -1136 & $>21.3$ & CFH12K     \\
20101113 & 55513.47 & -893  & $>20.8$ & CFH12K     \\
20101114 & 55514.51 & -892  & $>21.0$ & CFH12K     \\
20101115 & 55515.55 & -891  & $>21.0$ & CFH12K     \\
20101116 & 55516.46 & -890  & $>21.3$ & CFH12K     \\
20101117 & 55517.46 & -889  & $>21.1$ & CFH12K     \\
20101118 & 55518.46 & -888  & $>21.6$ & CFH12K     \\
20101119 & 55519.48 & -887  & $>19.5$ & CFH12K     \\
20101130 & 55530.44 & -876  & $>20.5$ & CFH12K     \\
20101201 & 55531.45 & -875  & $>22.0$ & CFH12K     \\
20101202 & 55532.53 & -874  & $>21.4$ & CFH12K     \\
20101205 & 55535.43 & -871  & $>20.5$ & CFH12K     \\
20101207 & 55537.38 & -869  & $>20.9$ & CFH12K     \\
20101208 & 55538.42 & -868  & $>21.4$ & CFH12K     \\
20101209 & 55539.50 & -867  & $>21.4$ & CFH12K     \\
20101211 & 55541.52 & -865  & $>21.2$ & CFH12K     \\
20101213 & 55543.53 & -863  & $>21.4$ & CFH12K     \\
20101231 & 55561.36 & -845  & $>20.9$ & CFH12K     \\
20110102 & 55563.39 & -843  & $>21.4$ & SI 600-277 \\
20110111 & 55572.27 & -834  & $>20.7$ & CFH12K     \\
20110112 & 55573.39 & -833  & $>21.3$ & CFH12K     \\
20110113 & 55574.43 & -832  & $>21.5$ & CFH12K     \\
20110114 & 55575.41 & -831  & $>21.4$ & CFH12K     \\
20110115 & 55576.43 & -830  & $>20.5$ & CFH12K     \\
20110116 & 55577.45 & -829  & $>20.6$ & CFH12K     \\
20110117 & 55578.45 & -828  & $>20.8$ & CFH12K     \\
20110118 & 55579.46 & -827  & $>20.9$ & CFH12K     \\
\hline
\end{tabular}
\end{minipage}
\end{table*}
\begin{table*}
\begin{minipage}{175mm}
\contcaption{}
\label{tab:continued}
\begin{tabular}{@{}cccc@{}}
\hline
20110119 & 55580.54 & $>$20.4 & CFH12K       \\
20110124 & 55585.37 & $>$20.7 & CFH12K       \\
20110125 & 55586.38 & $>$21.1 & CFH12K       \\
20110126 & 55587.46 & $>$20.8 & CFH12K       \\
20110128 & 55589.26 & $>$20.9 & CFH12K       \\
20110129 & 55590.33 & $>$21.5 & CFH12K          \\
20110130 & 55591.36 & $>$19.5 & CFH12K          \\
20110201 & 55593.43 & $>$21.3 & CFH12K          \\
20110209 & 55601.25 & $>$20.5 & CFH12K          \\
20110211 & 55603.25 & $>$22.1 & CFH12K          \\
20110212 & 55604.26 & $>$21.0 & CFH12K          \\
20110213 & 55605.28 & $>$21.1 & CFH12K          \\
20110214 & 55606.33 & $>$17.9 & CFH12K          \\
20110216 & 55608.26 & $>$20.9 & CFH12K          \\
20110222 & 55614.43 & $>$20.9 & CFH12K          \\
20110315 & 55635.15 & $>$20.9 & CFH12K          \\
20110417 & 55668.25 & $>$20.9 & CFH12K          \\
20120118 & 55944.36 & $>$21.1 & SI 600-277    \\
20120316 & 56002.29 & $>$20.5 & SI 600-386    \\
20120407 & 56024.01 & $>$20.7 & QUEST           \\
20120409 & 56026.05 & $>$21.3 & QUEST           \\
20120507 & 56055.00 & $>$21.6 & QUEST           \\
20121223 & 56284.48 & $>$20.2 & SI 600-386    \\
20130106 & 56298.29 & $>$20.2 & SI 600-277    \\
20130117 & 56309.46 & $>$19.7 & SI 600-277    \\
20130201 & 56324.35 & $>$20.9 & CFH12K          \\
20130202 & 56325.26 & $>$19.8 & SI 600-277    \\
20130204 & 56327.38 & $>$20.9 & CFH12K          \\
20130205 & 56328.27 & $>$21.6 & CFH12K          \\
20130206 & 56329.24 & $>$19.4 & CFH12K          \\
20130206 & 56329.36 & $>$19.8 & SI 600-277    \\
20130207 & 56330.28 & $>$21.5 & CFH12K          \\
20130301 & 56352.42 & $>$19.7 & SI 600-277    \\
20130303 & 56354.24 & $>$19.6 & CFH12K          \\
20130304 & 56355.28 & $>$21.3 & CFH12K          \\
20130305 & 56356.26 & $>$21.5 & CFH12K          \\
20130306 & 56357.23 & $>$21.3 & CFH12K          \\
20130311 & 56362.21 & $>$21.3 & CFH12K          \\
20130311 & 56362.31 & $>$20.3 & SI 600-386    \\
20130312 & 56363.26 & $>$21.5 & CFH12K          \\
20130313 & 56364.22 & $>$21.4 & CFH12K          \\
20130316 & 56367.17 & $>$21.6 & CFH12K          \\
20130317 & 56368.16 & $>$21.1 & CFH12K          \\
\hline
\end{tabular}

\medskip
SI 600-277: 0.7~m Schmidt telescope of the Catalina Sky Survey (CSS) with SI 600-277 \\
SI 300-386: 1.5~m Cassegrain reflector of the Mt.~Lemmon Survey (MLS) with SI 300-386 \\
CFH12K: 1.2~m Samuel Oschin Telescope of the Intermediate Palomar Transient Factory (iPTF) with CFH12K \\
QUEST: 60" ESO Schmidt Telescope of the La Silla Quest (LSQ) survey with QUEST. \\
\end{minipage}
\end{table*}
\onecolumn
\noindent
$^{1}$INAF - Osservatorio Astronomico di Padova, Vicolo dell'Osservatorio 5, 35122 Padova, Italy \\
$^{2}$Universit\`a degli Studi di Padova, Dipartimento di Fisica e Astronomia, Vicolo dell'Osservatorio 2, 35122 Padova, Italy \\
$^{3}$School of Physics and Astronomy, University of Southampton, Southampton, SO17 1BJ, UK \\
$^{4}$Department of Physics, Yale University, New Haven, CT 06520-8120, USA \\
$^{5}$Computational Cosmology Center, Computational Research Division, Lawrence Berkeley National Laboratory, 1 Cyclotron Road MS 50B-4206, \\ Berkeley, CA 94611, USA \\
$^{6}$Department of Astronomy, University of California, Berkeley, CA, 94720-3411, USA \\ 
$^{7}$Astronomy Department, California Institute of Technology, Pasadena, CA, 91125, USA \\   
$^{8}$Benoziyo Center for Astrophysics, Faculty of Physics, Weizmann Institute of Science, Rehovot 76100, Israel \\
$^{9}$Sternberg Astronomical Institute, M.V. Lomonosov Moscow State University, Universitetskii pr. 13, 119992 Moscow, Russia \\
$^{10}$Special Astrophysical Observatory, Nizhnij Arkhyz 369167, Russia \\ 
$^{11}$Kazan Federal University, Kazan, 420008, Russia \\
$^{12}$Millennium Institute of Astrophysics, Vicu\~{n}a Mackenna 4860, 7820436 Macul, Santiago, Chile \\
$^{13}$Instituto de Astrof\'isica, Facultad de F\'isica, Pontificia Universidad Cat\'olica de Chile, Vicu\~na Mackenna, 4860436 Macul, \\ Santiago de Chile, Chile \\
$^{14}$European Southern Observatory, Alonso de C\'ordova 3107, Vitacura, Casilla 19001, Santiago 19, Chile \\
$^{15}$Space Science Institute, 4750 Walnut Street, Suite 205, Boulder, Colorado 80301 \\
$^{16}$European Southern Observatories, Karl-Schwarzschild-Str., D-85748 Garching, Germany \\
$^{17}$Max-Planck-Institut f\"ur Astrophysik, Karl-Schwarzschild-Str. 1, D-85748 Garching, Germany \\
$^{18}$Las Cumbres Observatory Global Telescope Network, 6740 Cortona Dr., Suite 102, Goleta, CA 93117, USA \\
$^{19}$Department of Physics, University of California, Santa Barbara, Broida Hall, Mail Code 9530, Santa Barbara, CA 93106-9530, USA \\
$^{20}$Universidad de Atacama, Departamento de Fisica, Copayapu 485, Copiapo, Chile \\
$^{21}$Fundaci\'on Galileo Galilei - INAF, Telescopio Nazionale Galileo, Rambla Jos\'eAnaFern\'andez P\'erez 7, E-38712 Bre\~na Baja, Tenerife, Spain \\
$^{22}$Tuorla Observatory, Department of Physics and Astronomy, University of Turku, V\"ais\"al\"antie 20, FI-21500 Piikki\"o, Finland \\
$^{23}$Astrophysics Research Centre, School of Mathematics and Physics, Queen's University Belfast, Belfast BT7 1NN, UK \\ 
$^{24}$Astronomy/Physics MS HSB 314, One University Plaza Springfield, IL 62730, USA \\
$^{25}$Finnish Centre for Astronomy with ESO (FINCA), University of Turku, V\"ais\"al\"antie 20, FI- 21500 Piikki\"o, Finland \\
$^{26}$Institut de Ci\`encies de l'Espai (CSIC - IEEC), Campus UAB, Cam'i de Can Magrans S/N, 08193 Cerdanyola (Barcelona), Spain \\
$^{27}$Jet Propulsion Laboratory, California Institute of Technology \\
$^{28}$Centro Interdipartimentale Studi e Attivit\`a Spaziali (CISAS) -- G. Colombo, Universit\`a degli Studi di Padova,Via Venezia 15, 35131 Padova, Italy \\
$^{29}$Space and Remote Sensing, MS B244, Los Alamos National Laboratory, Los Alamos, NM 87545, USA \\
% Don't change these lines
\bsp	% typesetting comment
\label{lastpage}
\end{document}